\documentclass[12pt]{iopart}

\usepackage[pdftex]{graphicx}
\graphicspath{{figures/}}
\usepackage{subfig}

\usepackage{esint}
\makeatletter

\usepackage{amsbsy}

\makeatother

\begin{document}

\title{Cosmology in One Dimension: Fractal Geometry, Power Spectra and Correlation}

\author{Bruce N. Miller$^{1}$, Jean-Louis Rouet$^{2}$}

\address{$^{1}$Department of Physics and Astronomy, Texas Christian University,
Fort Worth, Texas 76129}

\address{$^{2}$Institut des Sciences de la Terre d'Orléans - UMR 6113 CNRS/Université
d'Orléans, OSUC, 1A, rue de la Férollerie, F--45071 Orléans Cedex
2}

\maketitle
\ead{ b.miller@tcu.edu}
\begin{abstract}
Concentrations of matter, such as galaxies and galactic clusters,
originated as very small density fluctuations in the early universe.
The existence of galaxy clusters and super-clusters suggests that a
natural scale for the matter distribution may not exist. A point of
controversy is whether the distribution is fractal and, if so, over
what range of scales. One-dimensional models demonstrate that the
important dynamics for cluster formation occur in the
position-velocity plane. Here the development of scaling behavior
and multifractal geometry is investigated for a family of
one-dimensional models for three different, scale-free, initial
conditions. The methodology employed includes: 1) The derivation of
explicit solutions for the gravitational potential and field for a
one-dimensional system with periodic boundary conditions (Ewald sums
for one dimension); 2) The development of a procedure for obtaining
scale-free initial conditions for the growing mode in phase space
for an arbitrary power-law index; 3) The evaluation of power
spectra, correlation functions, and generalized fractal dimensions
at different stages of the system evolution. It is shown that a
simple analytic representation of the power spectra captures the
main features of the evolution, including the correct time
dependence of the crossover from the linear to nonlinear regime and
the transition from regular to fractal geometry. A possible physical
mechanism for understanding the self-similar evolution is
introduced. It is shown that hierarchical cluster formation depends
both on the model and the initial power spectrum. Under special
circumstances a simple relation between the power spectrum,
correlation function, and correlation dimension in the highly
nonlinear regime is confirmed.
\end{abstract}

\section{Introduction}

The observation that visible matter in the universe shows structure
on a huge range of scales, from galaxies, to clusters, to super-clusters,
to voids \cite{peebles2}, has led Mandelbrot, Pietronero and others
to conjecture that the structure of the universe may be fractal \cite{cosfrac}.
Support for this controversial conjecture is provided by the fact
that the inter-galactic, two-body correlation function decays as a
power-law \cite{Martinez}. If the geometry of the universe was truly
fractal on all scales, then the basic principle of cosmology, that
the universe is homogeneous and isotropic on large scales, would be
violated. While current observations support the existence of an upper
bound for the size of the largest structures, the issue is not completely
closed \cite{bigvoid,Labini09}. Typically, in nature, observable
fractal behavior is usually restricted to a finite scale range \cite{Fed}.
Regardless of whether such a bound exists, it is still important to
understand how the appearance of fractal geometry develops from the
underlying dynamics.

The remarkable uniformity of the temperature distribution of the cosmic
background radiation CMB is a consequence of the nearly uniform energy
distribution at recombination \cite{WMAP5yr}. Over time, the small
(order $10^{-4}$) density fluctuations developed into the structures
we see today. Measurements and theory suggest that, following inflation,
the density fluctuations were Gaussian random variables with an approximate
power-law spectral density. They were subsequently modulated by baryonic
acoustic oscillation before recombination \cite{Martinez}. The commencement
of structure formation in the dark matter component actually preceded
recombination. These circumstances provide a natural scenario for
the initial conditions of any dynamical study. With current technology
astronomers can observe bright objects in the distant past. However,
they cannot directly observe dynamical processes that take giga-years
to unfold. As a consequence, computer simulation plays an especially
prominent role in astrophysics, and has been employed to investigate
complex behavior ranging from galaxy collisions and mergers to structure
formation\cite{Virgo}. Although progress has been substantial, the
ability of three-dimensional simulations with on the order of $10^{9}$
particles to resolve fractal scaling laws is still hampered by limited
resolution and approximations in the underlying dynamics. In contrast,
simulations with one-dimensional models can incorporate up to $10^{6}$
particles without compromising the two-body gravitational interaction.
Thus, although they are less realistic {}``toy'' models, they have
the potential to yield insights into clustering with current computers.
In addition, since theory suggests that the first collapsed objects
were highly flattened {}``pancakes'', they may yet provide contact
with the real universe \cite{peebles2,Zel}.

Rouet and Feix were the first to recognize the potential for using
one-dimensional models to investigate clustering in a matter-dominated
cosmological setting \cite{Rouet1,Rouet2}. Since there is no curvature
in one dimension, general relativity does not provide a unique path
for obtaining the correct laws of motion. Nonetheless, they showed
how the transformation to co-moving coordinates could be accomplished
in a completely self-consistent formulation which we refer to as the
RF (Rouet-Feix) model. Starting with a spatially uniform initial distribution,
they demonstrated that, as time evolved, hierarchical clustering occurs.
Thus the one-dimensional evolution is similar to what is believed
to have occurred in the universe following recombination. Rouet, Feix
and Jamin computed the box counting dimension for the distribution
of matter in the one-dimensional configuration and $\mu$ (position,
velocity) space. They obtained a result of about 0.6 for the configuration
space, strongly suggesting a fractal geometry. Since their seminal
work, another similar model was proposed by Fanelli and Aurell, referred
to simply as the quintic or Q model \cite{quintic}. As we will see,
the Q model sacrifices internal consistency in order to maintain the
correct coefficient of the average cosmological density.

A different approach employing a one-dimensional model to explore
matter-dominated cosmology was taken by Gouda and co-workers. Yano
and Gouda employed the Zeldovich approximation \cite{Zel} to investigate
the evolution of the one-dimensional system\cite{Gouda_powsp}. When
there is no cutoff in the initial spectra they showed that the evolution
of the power spectra is self-similar, and established that three different
scaling regimes occur, each with its own characteristic spectral index.
Later Yano \etal \cite{Gouda_caus} studied the development of a
single wave in the phase plane to investigate the effect of caustics
on the evolution of the power spectra. In addition, Tatekawa and Meida
used the Zeldovich approximation to study the self-similar evolution
of a one-dimensional system with initial conditions selected from
a Cantor set\cite{Tat}.

More recently Miller and Rouet extended the original work of Rouet
and Feix to include an investigation of multi-fractal properties \cite{MRexp,MRGexp}.
They considered the fact that the clusters are actually forming in
$\mu$ space. An analysis of the generalized fractal dimensions was
applied to both configuration and $\mu$ space and it was shown that
multifractal geometry develops in each. Miller and Rouet carried out
a multifractal analysis of both the RF and Q models, as well as a
Hamiltonian version. Valageas has investigated the Hamiltonian version
with identical boundary conditions, and has shown the existence of
a sequence of stationary and equilibrium states\cite{Val1,Val2}.
The use of one-dimensional models has recently been extended by Sutter
and Ricker to include a dark energy field in addition to dark matter
\cite{Ricker1d}. They investigated how the introduction of dark energy
influences the formation of Zeldovich {}``pancakes'', presumably
the first large-scale structures to emerge in the cosmos. In a recent
series of articles Gabrielli \etal have studied the one-dimensional
system in the infinite particle limit \cite{gab_pp,Gab,Joyce_1d,Gab_Joy_1DPD,Joyce_range}.
A central focus of the series is the statistical properties of the
force distribution obtained from sampling static distributions of
particle positions from infinite perturbed lattices. In \cite{Joyce_1d}
a number of the issues mentioned here are discussed within this framework.

Theoretical cosmology suggests that during the period that density
fluctuations remain small, evolution is linear and is dominated by
a {}``growing mode'' and the spectral distribution of the fluctuations
remains a power-law \cite{peebles2}. However, in the simulations
of Miller and Rouet described above, initial conditions were created
by independently sampling the velocity of each particle from either
a uniform or Gaussian distribution. In the work presented here, the
initial conditions follow the cosmological picture for cold dark matter.
We have performed simulations and a multi-fractal analysis of each
model for a variety of initial, power-law, spectral densities. Here
we describe the circumstances under which the evolution follows the
accepted pattern of self-similar, hierarchical clustering. We show
how the evolution of the spectral density and correlation function
depend on both the model and the initial power-law index. We find
that, for sufficiently large samples, the analytic relation between
correlation dimension, correlation function, and spectral density
is obeyed.

In the following we first explain the mathematical formulation of
the different models in section \ref{1D}. In particular, we show
how to construct the exact gravitational potential and field for a
one-dimensional, periodic system, \textit{i.e.}, Ewald sums for one
dimension. In Section \ref{Simulation-IC} we describe how simulations
are performed and show how to create scale-free initial conditions
for the growing mode in phase space for an arbitrary initial scaling
index. We also present typical results for the Q model in the highly
nonlinear region, including the calculation of power spectra and the
two-body correlation function. In addition a simple analytic representation
of the power spectra is introduced that captures the main features
of the evolution, including the correct time dependence of the crossover
from the linear to nonlinear regime and the transition from regular
to fractal geometry. In Section \ref{fractal-measure} we introduce
techniques for measuring generalized fractal dimensions and apply
them to the Q model. In addition we give a comparison of results for
related models and different initial power spectra and demonstrate
that the power spectra, correlation function, and correlation dimension
are interrelated in the nonlinear regime. To provide a physical basis
for the observed behavior we introduce the mechanism of hierarchical
virialization. Finally, in the last section, we present general conclusions
that can be drawn from this investigation and provide a discussion
of some open questions that still need to be addressed.

\section{One-Dimensional Systems}

\label{1D}

Although dynamics in the universe is governed by the general theory
of relativity, in a sufficiently small sample Newtonian dynamics provides
an adequate working model \cite{Newtap}. In order to study structure
formation, cosmologists focus on a segment of the universe $\Omega$
that is small enough that a Newtonian description is adequate, but
much larger than the two-body correlation length \cite{Virgo}. They
represent the location of the mass points of N-body simulations in
comoving coordinates that follow the Hubble flow, so the average density
remains constant. They ascribe a cubical shape to the sample and assert
periodic boundary conditions to obtain a closed dynamical system that
mimics a segment of the real universe \cite{Bert_rev}. This conserves
mass and insures the continuity and smoothness of the gravitational
potential and field at the boundaries. A potential problem is that,
on the average, the universe is isotropic but the symmetry imposed
by the cubical boundary conditions is not. For a large enough cube,
this problem can be avoided for a finite time. In tree models Ewald
sums are typically employed to compute the gravitational force from
the infinite number of system {}``copies'' \cite{hern_ewald} while
in grid models the periodic boundary conditions are built into the
potential \cite{HockneyEastwood}.

A one-dimensional gravitational system corresponds to a set of parallel
mass sheets moving in the direction perpendicular to their surface.
In the seminal work of Rouet and Feix, they assumed a universe with
spherical symmetry about a point. The elements of their universe were
then concentric, spherical, irrotational mass shells. Far from the
symmetry center, the radius of curvature is large and, locally, the
shells are approximately planar and parallel. By choosing such a segment,
they obtained a stratified system of comoving, planar, mass sheets
for their model. However, if the object is to construct a consistent,
one-dimensional, gravitational model that embraces an expanding background,
it is not necessary to assume any special symmetry. One can just start
with one dimension and assert a uniform expansion factor that also
applies to the parallel dimensions of the mass planes. If gravity
is the only force acting, then internal consistency forces the $t^{2/3}$
time dependence associated with the expansion factor of the Einstein
de-Sitter, or matter-dominated, universe. However, the coefficient
of the time-dependent mean density differs from the regular, three-dimensional
model. To avoid this mild conundrum Fanelli and Aurell took a different
approach. They first transformed to comoving coordinates in homogenous
and isotropic three-space and then inserted a system of planes. Because
the negative background density induced by the transformation to comoving
coordinates doesn't cancel with the average density associated with
the mass sheets, an additional correction has to be included \cite{MRGexp}.
Here we will follow this approach.

Consider such a bounded region $\Omega$. We are interested in the
evolution of density fluctuations following the time of recombination,
so that electromagnetic forces can be ignored and Newtonian dynamics
provides an adequate representation of the motion in a finite region
\cite{Newtap}. Then, in a (3+1)--dimensional universe, the Newtonian
equations governing a mass point are simply

\begin{equation}
\frac{d\mathbf{r}}{dt}=\mathbf{v,\qquad}\frac{d\mathbf{v}}{dt}=\mathbf{E}_{g}\mathbf{(r,}t\mathbf{)}\label{3d}\end{equation}
 where, here, $\mathbf{E}_{g}\mathbf{(r,}t\mathbf{)\ }$ is the gravitational
field. To follow the motion in a frame of reference where the average
density remains constant, \textit{i.e.} the comoving frame, we introduce
the scale factor $A(t)$ for a matter-dominated universe \cite{peebles2}
and transform to a new space coordinate which scales the distance
according to $A(t)$. Writing $\mathbf{r}=A(t)\mathbf{x}$ we obtain

\begin{equation}
\frac{d^{2}\mathbf{x}}{dt^{2}}+\frac{2}{A}\frac{dA}{dt}\frac{d\mathbf{x}}{dt}+\frac{1}{A}\frac{d^{2}A}{dt^{2}}\mathbf{x=}\frac{1}{A^{3}}\mathbf{E}_{g}(\mathbf{x},t)\label{scaldis}\end{equation}
 where, in the above, we have taken advantage of the inverse square
dependence of the gravitational field to write $\mathbf{E}_{g}(\mathbf{x},t)=\frac{1}{A^{2}}\mathbf{E}_{g}\mathbf{(r},t)$
where the functional dependence is preserved. In a matter-dominated
(Einstein-de Sitter) universe we find that

\begin{equation}
A(t)=\left(\frac{t}{t_{0}}\right)^{\frac{2}{3}},\qquad\rho_{b}(t)=\left(6\pi Gt^{2}\right)^{-1}\label{scale factor}\end{equation}
 where $t_{0}$ is some arbitrary initial time corresponding, say,
to the epoch of recombination, $G$ is the universal gravitational
constant, and $\rho_{b}(t)$ is the average, uniform, density frequently
referred to as the background density. The justification for equation
(\ref{scale factor}) comes from the Robertson-Walker metric and the
Friedman equation \cite{peebles2}. However, these results can also
be obtained directly from equation (\ref{scaldis}) by noting that
if the density is uniform so that all matter is moving with the Hubble
flow, the first two terms in equation (\ref{scaldis}) vanish whereas
the third term (times $A$) must be equated to the gravitational field
resulting from the uniformly distributed mass contained within a sphere
of radius $A\left|\mathbf{x}\right|$ which is simply the right-hand
side of equation (\ref{scaldis}). Then the third term of equation
(\ref{scaldis}) is the contribution arising by subtracting the field
due to the background density from the uniform sphere \cite{peebles2}.
Noting that $A^{3}\rho_{b}(t)=\rho_{b}(t_{0})$ forces the result.
Alternatively, also for the case of uniform density, taking the divergence
of each side of equation (\ref{scaldis}) and asserting the Poisson
equation forces the same result. Thus the Friedman scaling is consistent
with the coupling of a uniform Hubble flow with Newtonian dynamics
\cite{peebles2}.

In standard three-dimensional cosmological simulations it is common,
but not ubiquitous, to employ conformal time, i.e to use A(t) as the
measure of progress \cite{HockneyEastwood}. Here, for computational
purposes, we will see that it is useful to obtain autonomous equations
of motion with coefficients that do not depend explicitly on the time.
This can be effectively accomplished \cite{Rouet1,Rouet2} by transforming
the time coordinate according to

\begin{equation}
dt=B(t)d\tau,\qquad B(t)=\frac{t}{t_{0}}\label{timescal}\end{equation}
 yielding the autonomous equations

\begin{equation}
\frac{d^{2}\mathbf{x}}{d\tau^{2}}+\frac{1}{3t_{0}}\frac{d\mathbf{x}}{d\tau}-\frac{2}{9t_{0}^{2}}\mathbf{x=E}_{g}(\mathbf{x}).\label{auto}\end{equation}
 These can be further simplified by choosing the inverse Jeans' frequency
$T_{J}$ for the unit of time \cite{b&t}

\begin{equation}
T_{J}=\omega_{J}^{-1}=(4\pi G\rho)^{-1/2}=\sqrt{\frac{3}{2}}t_{o}\label{eq:JF}\end{equation}
 yielding

\begin{equation}
\frac{d^{2}\mathbf{x}}{d\tau^{2}}+\frac{1}{\sqrt{6}}\frac{d\mathbf{x}}{d\tau}-\frac{1}{3}\mathbf{x=E}_{g}(\mathbf{x)}\label{eq:fin3D}\end{equation}
 where we used the fact that $3t_{0}^{2}/2=1$ in the adopted time
units and $\tau$ is now expressed in the new units. Thus equation
(\ref{auto}) corresponds to a dissipative dynamical system in the
comoving frame with friction constant $1/\sqrt{6}$ and with forces
arising from fluctuations in the local density with respect to a uniform,
three-dimensional, isotropic, neutralizing background.

We now imagine that the actual source of the density fluctuations
is a system of parallel mass sheets with a neutralizing background,
similar to a single-component plasma \cite{lieb_matt,mattis}. For
the special case of the stratified mass distribution induced by the
one-dimensional system, the local particle density at time $t$ is
given by

\begin{equation}
\rho(x,t)=\sum m_{j}(t)\delta(x-x_{j})\label{den}\end{equation}
 where $m_{j}(t)$ is the mass per unit area of the $j^{th}$ sheet
and, from symmetry, the gravitational field only has a component in
the $x$ direction. To avoid confusion, we will just refer to this
as the mass. From Gauss' law we know that the field due to a single
mass sheet is constant. Then, for an isolated system with a finite
number of particles and no background, the field experienced by one
of the sheets is proportional to the difference between the mass on
each side \cite{YM2ma}. To obtain equations of motion, if we just
take the component of equation (\ref{eq:fin3D}) in the $x$ direction
we see that a problem arises. The third term corresponds to the field
from a sphere of constant density, but we now have a slab geometry
so this term needs to be multiplied by a factor of three to maintain
mass neutrality \cite{MRGexp}.

Following the three dimensional example, we will also assume periodic
boundary conditions to approximate the actual behavior in a finite
slab of our model one-dimensional universe for a finite time. The
size of the system will be determined by the length of elapsed time
we require to evolve the system before the presence of boundaries
play a significant role. To accomplish this we need to determine the
gravitational field induced by a single particle in the periodic system.
Assume that the width, and hence the spatial period, of our system
in the comoving frame is $2L,$ so we may choose the position $x_{1}$
of our particle of mass $m_{1}$ in $[-L,\: L)$ with the points at
$L$ and $-L$ identified. For periodic boundary conditions the potential
can only be defined for a mass-neutral system \cite{HockneyEastwood,Bert_rev,hern_ewald}
where the source of the potential and field is the difference between
the local density and its average over one period. Since our particle
carries with it the negative background density $-m_{1}/2L$, the
potential it induces, $\phi_{1}(x)$, satisfies the following Poisson
equation
\begin{equation}
\frac{\partial^{2}\phi_{1}(x)}{\partial x^{2}}=4\pi m_{1}G[\delta(x-x_{1})-\frac{1}{2L}]\label{eq:poiseq}
\end{equation}
 with the general solution
 \begin{equation}
\phi_{1}(x)=2\pi m_{1}G[\left|x-x_{1}\right|-\frac{1}{2L}(x-x_{1})^{2}+b(x-x_{1})+c]\label{eq:1psol}
\end{equation}
 where $b$ and $c$ are constants.

The periodicity requirement, $\phi_{1}(L)=\phi_{1}(-L)$, forces $b=0$.
In addition, to guarantee that as $L\rightarrow\infty$ the potential
approaches that of the isolated system, we choose the arbitrary additive
constant $c=0$, yielding for the final form of the potential\begin{equation}
\phi_{1}(x)=2\pi m_{1}G[\left|x-x_{1}\right|-\frac{1}{2L}(x-x_{1})^{2}].\label{eq:1psolf}\end{equation}
 We immediately obtain the gravitational field $E_{1}$ induced by
a single particle in the periodic system: \begin{equation}
E_{1}(x)=-\frac{\partial\phi_{1}(x)}{\partial x}=2\pi m_{1}G\left[\frac{1}{L}(x-x_{1})+\Theta(x_{1}-x)-\Theta(x-x_{1})\right]\label{eq:1pfield}\end{equation}
 where $\Theta(x)$ is the usual step function and, for consistency,
we assign $\Theta(0)=\frac{1}{2}.$ Note that we could have obtained
the result from symmetry as well. Periodic boundary conditions in
one dimension restrict the motion to the one-torus or circle. Since
there is no preferred direction on the circle, the spatially averaged
field must vanish. Thus symmetry alone guarantees that $b=0$. The
results for the potential and field, equations (\ref{eq:1psolf},\ref{eq:1pfield})
can also be obtained by employing a screening function to directly
sum over the periodic system images, or replicas \cite{MR_per}, and
are effectively Ewald sums for one dimension. Unlike their three-dimensional
analogs \cite{hern_ewald,brush}, they have a simple closed-form expression.

Let us assume that the primitive cell of our periodic system contains
$2N$ particles (sheets) confined within a slab with width $2L$,
\textit{i.e.} $-L\leq x<L$. For the special case of equal masses
$m_{j}(t)=m(t)$, by direct summation the correct form of the gravitational
field occurring at the location of particle $i$ in comoving coordinates
is then

\begin{equation}
E_{g}(x_{i})=2\pi m(t_{0})G[\frac{2N}{L}(x_{i}-x_{c})+N_{R,i}-N_{L,i}]\label{field}\end{equation}
 since we already implicitly accounted for the fact that $m_{j}(t)=m(t_{0})/A^{2}$
in equation (\ref{auto}). In equation (\ref{field}) $N_{R,i}$ ($N_{L,i}$)
is the number of sheets on the right (left) of particle $i$ and $x_{c}(t)$
is the system center of mass in $[-L,\: L)$. The field is further
simplified by establishing the connection between $m(t_{0})$ and
the background density at the initial time, $\rho_{b}(t_{0}).$ Then

\begin{equation}
\rho_{b}(t_{0})=\left(6\pi Gt_{0}^{2}\right)^{-1}=\left(\frac{N}{L}\right)m(t_{0}),\end{equation}
 and we may express the field by

\begin{equation}
E_{g}(x_{i})=\frac{1}{3t_{0}^{2}}\left(\frac{L}{N}\right)[\frac{2N}{L}(x_{i}-x_{c})+N_{R,i}-N_{L,i}].\end{equation}
 The equations of motion for a particle in the system now read

\begin{equation}
\frac{d^{2}x_{i}}{d\tau^{2}}+\frac{1}{\sqrt{6}}\frac{dx_{i}}{d\tau}\mathbf{=}\left(\frac{1}{2}\right)[2(x_{i}-x_{c}(\tau))+N_{R,i}-N_{L,i}].\label{eq:final}\end{equation}
 where we have taken account of the adopted unit of time (see equation
(\ref{eq:JF})) and chosen the mean inter-particle spacing $L/N$
for the unit of length.

The description is completed by noting that, since the system satisfies
periodic boundary conditions on the interval $[-L,\: L)$, when a
particle leaves the primitive cell defined by $-L\leq x<L$ on the
right (left), it re-enters at the left (right) hand boundary with
the identical velocity. Note that from equation (\ref{field}) there
is no change in the field experienced by other particles during such
a boundary crossing. This is a necessary condition for the one-torus
geometry, since there is nothing special about this point. Other consequences
of the torus geometry are discussed in \cite{MR_per}. In our simulations
we made the further assumption of symmetry about the origin: For each
particle with $0<x<L$ with velocity $v$, there is a twin located
at $-x$ with velocity $-v$. With this stipulation, $x_{c}=0$ in
equation (\ref{eq:final}) and the periodic boundary conditions are
equivalent to an $N$-particle system with reflecting boundary conditions
at $x=0,\, L$. For the times of interest, we demonstrated in \cite{MR_per}
that the behavior of the system is nearly identical for each type
of boundary. Note that, without the symmetry requirement, if one arbitrarily
sets $x_{c}=0$ the torus geometry is violated.

As we mentioned earlier, by embedding the stratified system into a
region of three-dimensional Euclidean space we are changing the local
symmetry, so we have to adjust the local background density. In concept
it is similar to the well-known Zeldovich approximation that is used
to investigate the formation of the first matter concentrations that
are believed to have a {}``pancake'' geometry \cite{peebles2}.
In their investigation of the structure of Zeldovich pancakes Aurell
\etal have shown that the models are closely related \cite{Aurell}.
In \cite{Joyce_1d} it is argued that they are equivalent. Notice
that this hybrid model mixes two different symmetries, the isotropic
3+1 dimensional system that cosmologists model with a periodic cube,
and the planar geometry that is represented by a slab. Since the mass
concentration in the idealized planes stretches to infinity in two
directions it cannot represent the physical universe. The slab geometry
can only represent cosmological evolution for a short time before
the influence of the other dimensions (terms in the strain tensor)
become important \cite{peebles2}. In contrast, the RF model is obtained
from the reverse sequence where one first restricts the geometry to
1+1 dimensions and then introduces the transformation to the comoving
frame. In this approach the derivation is more straightforward; it
is not necessary to make the adjustment in the coefficient of $x_{i}$
as we did here to obtain the correct background contribution \cite{Rouet1,Rouet2,MRexp}.
This is quickly seen by noting that the divergence of $\mathbf{x}$
is three times greater than the divergence of $x\boldsymbol{\hat{x}}$
which one would obtain by directly starting with the one-dimensional
model. As a consequence, in the RF model, the coefficient of the first
derivative term (the friction constant) in equation (\ref{eq:final})
is $1/\sqrt{2}$ instead of $1/\sqrt{6}$. In every other respect
the models are identical. This simply illustrates that, since there
is no curvature in a (1+1)-dimensional universe, there is a degree
of arbitrariness in choosing the final model. It cannot be obtained
solely from general relativity. For a discussion of this point, see
Mann \etal \cite{Mann1}. A Hamiltonian version can also be considered
by setting the friction constant equal to zero. In their earlier work,
using the linearized Vlasov-Poisson equations, Rouet and Feix carried
out a stability analysis of the model without friction. They determined
that the system followed the expected behavior, \textit{i.e.}, when
the system size is greater than the Jeans' length, instability occurs
and clustering becomes possible \cite{Rouet1,Rouet2}. Of course,
when the friction term is not present, both the Q and RF models are
identical. Then, with the assumption that the friction term will not
have a large influence on short-time linear stability, the analysis
of the Hamiltonian version applies equally to both versions.

\section{Simulations and Initial Conditions}

\label{Simulation-IC}

An appeal of these one-dimensional gravitational systems is their
ease of simulation. In each of the one-dimensional Newtonian models
considered here (Q, RF and Hamiltonian), it is possible to analytically
integrate the motion of the individual particles between crossings.
Then the temporal evolution of the system can be obtained by following
the successive crossings of the individual, adjacent, particle trajectories.
In particular, for the Q model, if we let $y_{i}=x_{i+1}-x_{i}$,
where we have ordered the particle labels from left to right, then
we find that the differential equation for each $y_{i}$ is the same,
namely

\begin{equation}
\frac{d^{2}y_{i}}{d\tau^{2}}+\frac{1}{\sqrt{6}}\frac{dy_{i}}{d\tau}-y_{i}=-1\label{disp}\end{equation}

The general solution of the homogeneous version of equation (\ref{disp})
is a sum of exponentials. By including the particular solution of
the inhomogeneous equation (simply a constant) we obtain a fifth order
algebraic equation in $u=\exp(\tau/\sqrt{6})$ for the successive
crossings, defined by $y_{j}(\tau)=0$ ; hence the name Q, or quintic,
model. These can be determined numerically in terms of the initial
conditions by analytically bounding the roots and employing a numerical
root-finding method. Note that for the RF model a cubic equation is
obtained so the crossing times can be found analytically \cite{Rouet1,Rouet2,MRexp}.
A sophisticated, event-driven algorithm was designed to execute the
simulations. Using the Newton-Raphson method, the algorithm computes
all possible crossings of adjacent pairs of particles with double-precision
accuracy. Since, between crossings, the general solution of the dynamical
equations is known analytically, once a crossing time is established,
the position and velocity of each particle can be determined with
the accuracy of the computer. Two important features of the algorithm
are that it only updates the positions and velocities of a pair of
particles when they actually cross, and it maintains the correct ordering
of each particle's position on the line. In contrast with typical
N-body simulations, it is not necessary to introduce a discreteness
length or time step. Using the algorithm we are able to carry out
runs for significant cosmological times with large numbers of particles.
In particular, it is possible to simulate a system until all of the
mass is concentrated in just a few clusters. Depending on the initial
conditions, this typically occurs on the order of 15-20 dimensionless
time units into the simulation. Since, at this stage, the influence
of the boundary conditions can no longer be ignored, there is no advantage
in continuing the runs any further.

In cosmology, a consequence of inflation is that primordial density
fluctuations are independent Gaussian random variables with a scale-free,
power-law, spectral power distribution $P(k)\sim k^{n}$ \cite{peebles2}.
Analysis of the WMAP measurements of the angular distribution of the
CMB \cite{WMAP5yr} yields a value for $n_{s}$ very close to unity,
the Harrison-Zeldovich value. In this study we have investigated the
dynamics of all three models with three different initial power spectra
with indices, $n=0,\,1,\,2$. To construct the initial conditions
we assume that the system is nearly uniform. If we use an ordered
labeling of the $N$ particles, then the equilibrium positions are
simply $x_{j0}=\frac{L}{N}(j-\frac{1}{2})$ for $j=1,...,N.$ Initially
the system is in the linear regime and no crossings have occurred
so the particles are assumed to be very close to their equilibrium
points. Then, from equation (\ref{field}) it is straightforward to
show that the gravitational field on particle $j$ with position $x_{j}$
arising from the initial density fluctuation is, in our units, simply
$\delta E(x_{j})=x_{j}-x_{j0}\equiv z_{j}$. Represent the local density
fluctuation $\delta\rho(x)$ as a Fourier series with coefficients
$\delta\rho_{k}$. Then, from the Poisson equation, \[
\delta E(x_{j})=-4\pi G\sum_{k}\frac{1}{ik}\delta\rho_{k}(e^{ikx_{j}}-1)\]
 where we note that, at the origin, $\delta E$ vanishes and there
is no contribution from $k=0$. Taking the Fourier coefficients as
Gaussian random variables with a power-law power spectrum $\left|\delta\rho_{k}\right|^{2}$
of index $n$, we have $\delta\rho_{k}\sim k^{n/2}e^{i\theta_{k}}$
where the random phases $\theta_{k}$ are uniformly distributed on
$\left[0,\,2\pi\right]$ and $\theta_{-k}$=$-\theta_{k}$ to insure
$\delta\rho$ is real. Combining the above we obtain \[
z_{j}=x_{j}-x_{j0}=C\sum_{k>0}k^{n/2-1}[sin(kx_{j0}+\theta_{k})-sin(\theta_{k})],\quad\: k=\upsilon\pi/N,\:\upsilon=1,...,N\]
 for the initial displacements of the particles from equilibrium.
The sum is cut off when the wavelength is smaller than the mean inter-particle
spacing. The constant of proportionality $C$ was chosen just small
enough to prevent any crossings of the ordered positions. In the linear
regime, from equation (\ref{eq:final}), the evolution of each particle
is governed by \begin{equation}
\frac{d^{2}z_{i}}{d\tau^{2}}+\frac{1}{\sqrt{6}}\frac{dz_{i}}{d\tau}=\delta E(x_{j})=z_{j}\label{eq:linreg}\end{equation}
 with general solution \[
z_{j}(\tau)=a\exp\left(\frac{2}{\sqrt{6}}\tau\right)+b\exp\left(-\frac{3}{\sqrt{6}}\tau\right).\]
 Following standard practice, we include only the contribution from
the growing mode. This fixes the initial velocity of particle $j$
as $\frac{2}{\sqrt{6}}z_{j}$ and completes the assignment of the
initial condition of the N-body system. For an alternative approach
for indices $n=0,1,2$ see \cite{gab_pp}.

\begin{figure}[ht]
 \centerline{\includegraphics[width=1\textwidth]{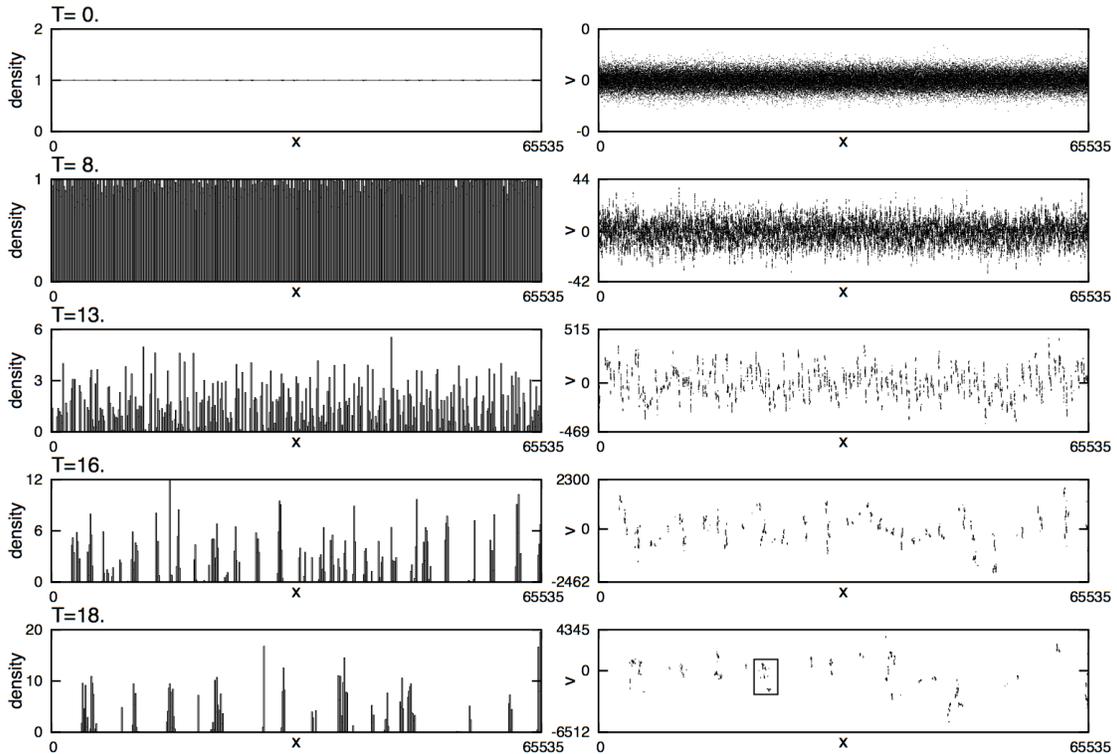}}
\caption{\label{snap} Evolution in configuration and $\mu$ space for the
quintic model with $2^{16}$ particles from $\tau=0$ to $\tau=18$.
The initial distribution is such that the density power spectrum has
a power-law of index $n=2$ (see Figure \ref{inspec}).}

\end{figure}

\begin{figure}[ht]
 \centerline{ \subfloat{(a)}{\includegraphics[height=0.25\textwidth]{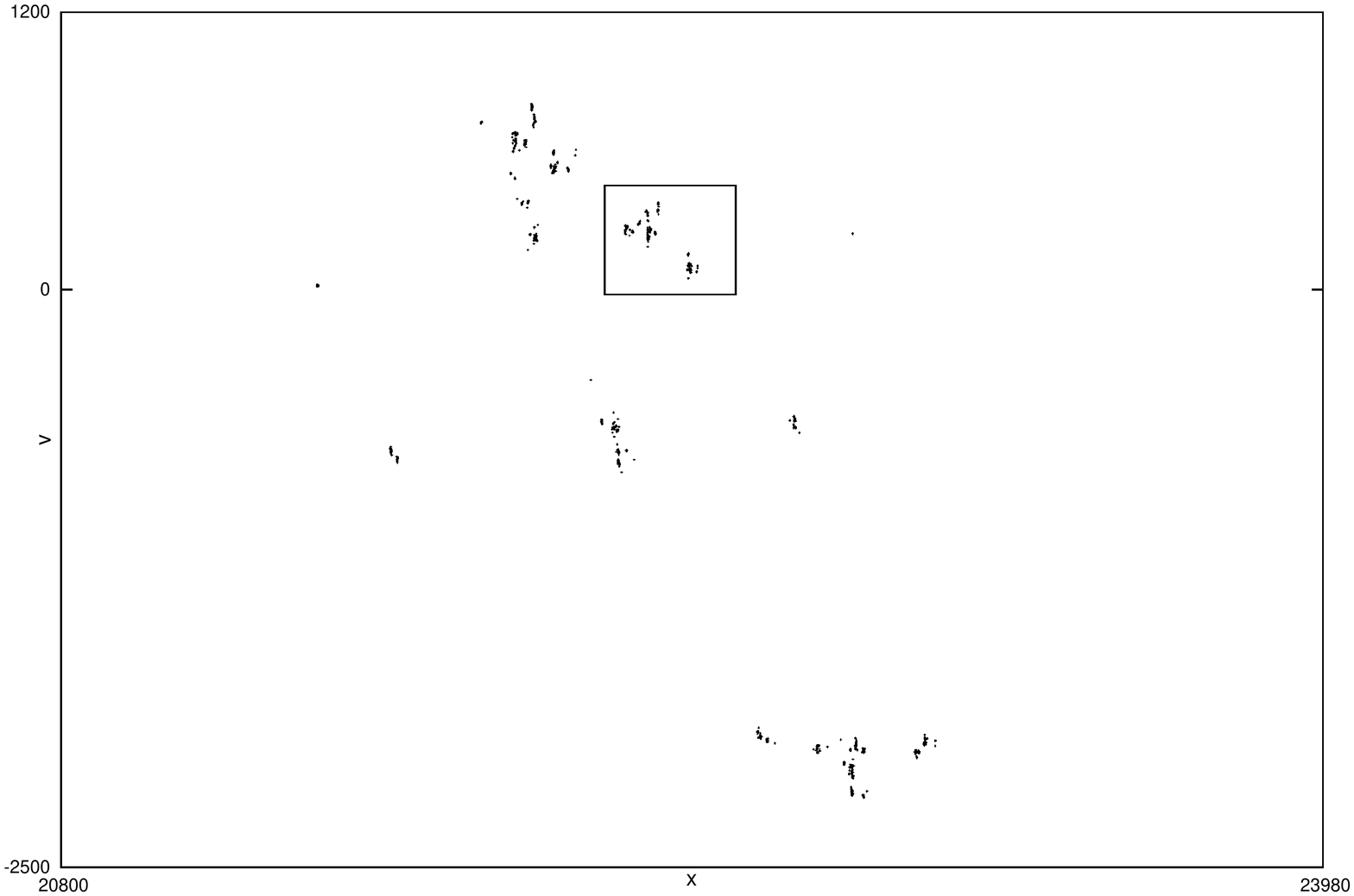}}
\subfloat{(b)}{\includegraphics[height=0.25\textwidth]{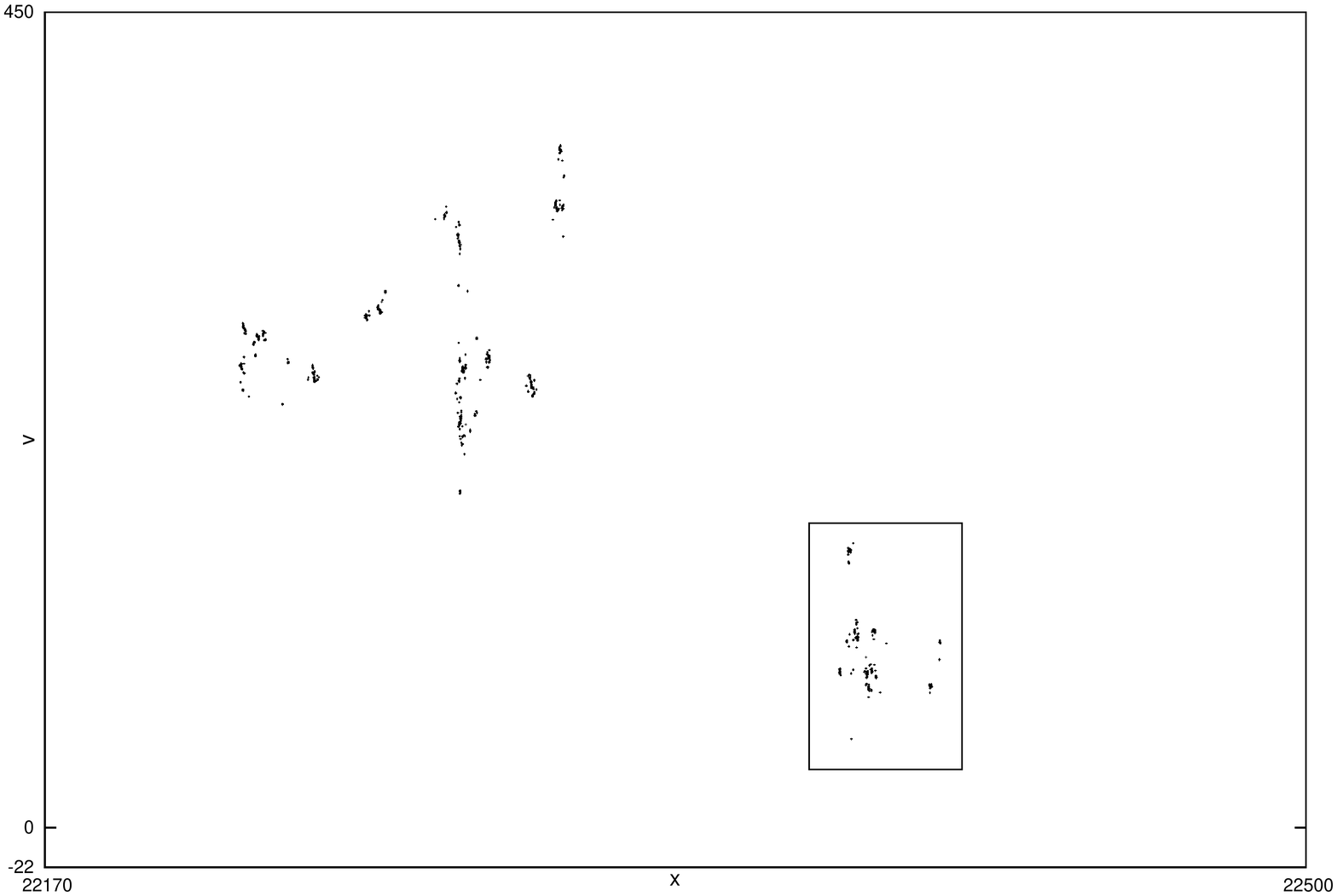}}
\subfloat{(c)}{\includegraphics[height=0.25\textwidth]{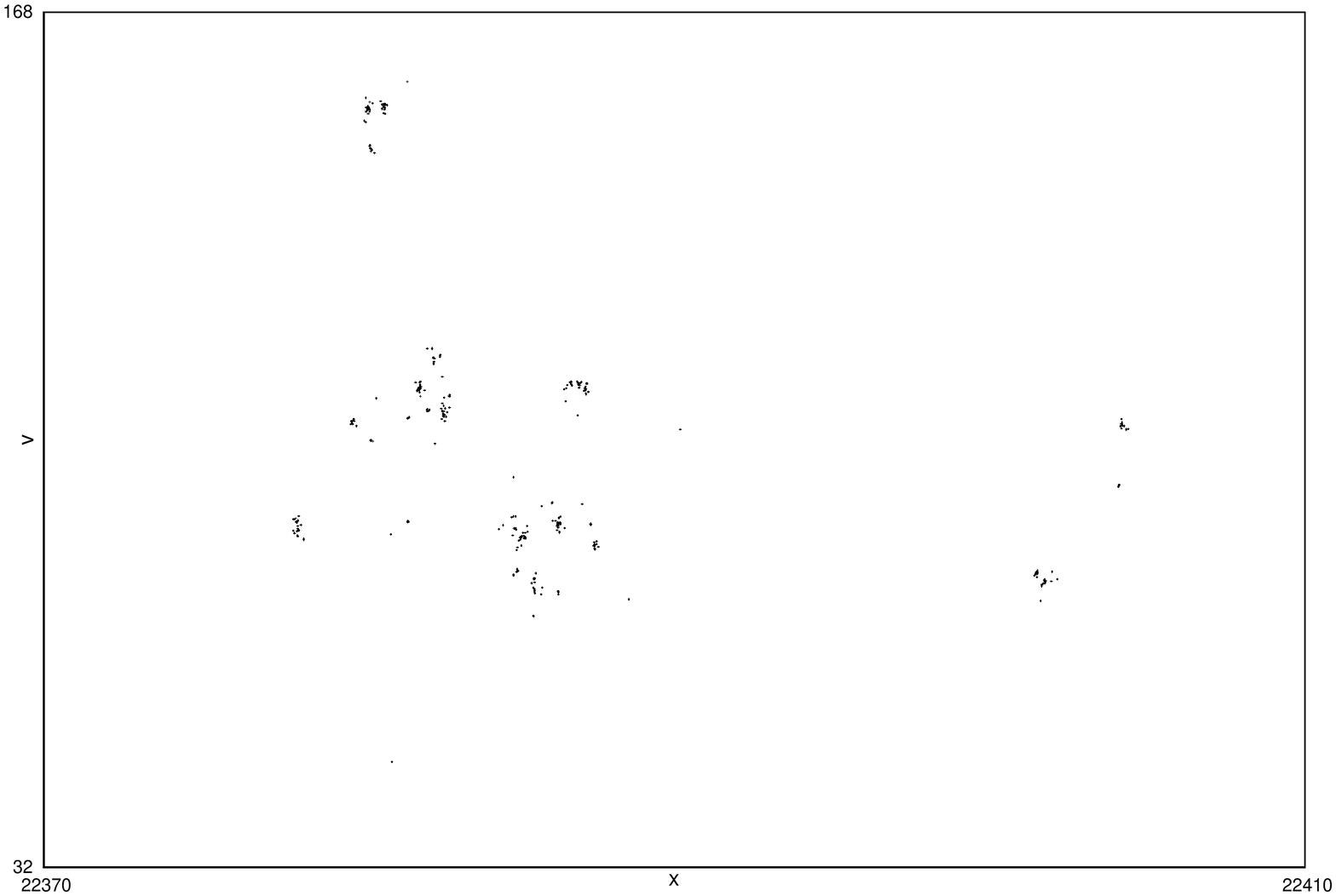}}}
\caption{\label{zoom0-zoom2} Consecutive expansions (zooms) in the $\mu$
space panels. Panel (a) shows the region selected by a rectangle in
the $\mu$ space panels at $\tau=18$ of figure \ref{snap}. Panel
(b) is the region selected by a rectangle in panel (a) and so on.
These representations have the appearance of a random fractal which
suggests self-similarity.}

\end{figure}

In figure \ref{snap} we present a visualization of a typical run
with $2^{16}$ particles. The system consists of the Q model with
an initial spectral index of $n=2$. Initially the velocity spread
is small, within (-1, 1) in the dimensionless units employed here,
and the system contains on the order of $10^{5}$ Jeans' lengths.
Of course, since the initial state does not represent thermal equilibrium,
the Jeans' length lacks predictive certainty, but strongly suggests
that instability will ensue. Following the original work of Rouet
\etal \cite{Rouet1,Rouet2}, we present a sequence of snapshots at
integer values of the dimensionless scaled time through $\tau=18$.
In the left column we present a histogram of the particle positions
at increasing time frames, while on the right we display the corresponding
particle locations in $\mu$ space. It is clear from the panels that
hierarchical clustering is occurring, \textit{i.e.}, small clusters
are joining together to form larger ones, so the clustering mechanism
is {}``bottom-up''\cite{peebles2}. The first clusters seem to appear
at about $\tau=8$ and there are many, while by $\tau=18$ there are
only on the order of a dozen clusters. In larger simulations in the
$\mu$ space we observe that, between the clusters, matter is distributed
along linear paths \cite{MRGexp}. As time progresses the size of
the linear segments bridging the clusters increases. The behavior
of these under-dense regions is governed by the stretching in $\mu$
space predicted analytically by Vlasov theory \cite{MRexp,MRGexp}.
Qualitatively similar histories are obtained for the RF model but
the clustering occurs more slowly. However, there are some subtle
differences. In figures \ref{zoom0-zoom2} we zoom in and show magnified
inserts from the mass distribution in $\mu$ space at time $\tau=18$.
The hierarchical structure observed in these models suggests the existence
of fractal geometry, but careful analysis is required to determine
if this is correct.

\begin{figure}[ht]
 \centerline{\includegraphics[width=1\textwidth]{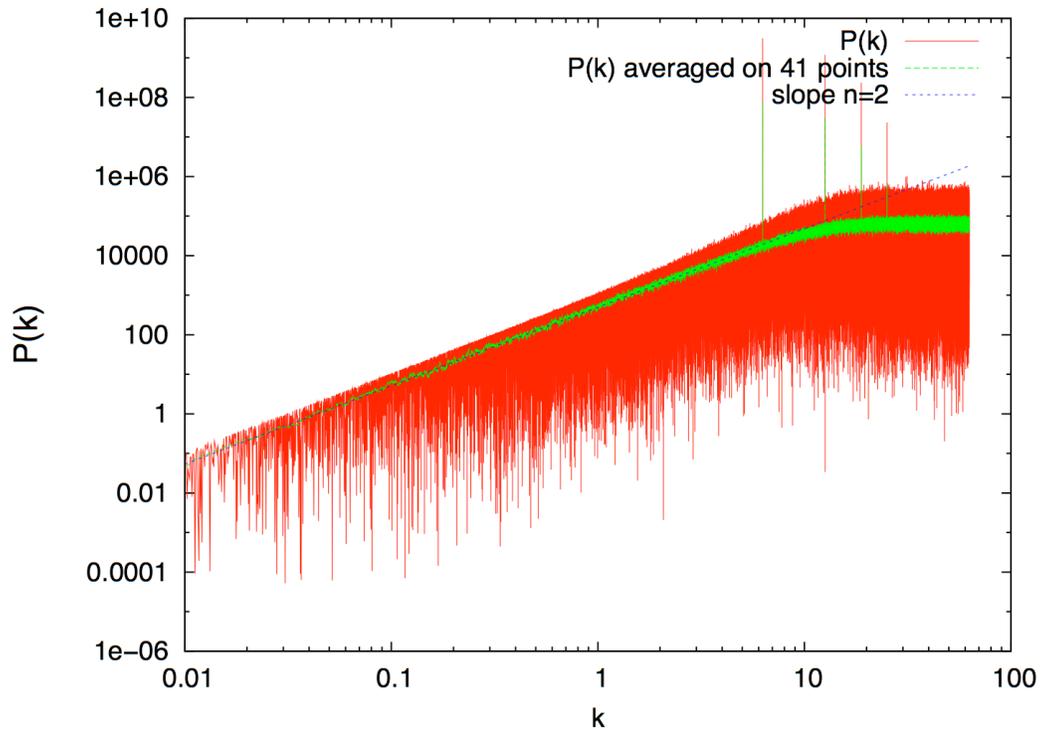}}
\caption{\label{inspec} The Power spectrum of the initial density fluctuations
for the quintic model. Particles have been distributed so that the
power spectrum presents a power-law of index $n=2$.}

\end{figure}

In figure \ref{inspec} we show the power spectra of the initial density
fluctuations $\left|\delta\rho_{k}\right|^{2}$. Although the data
are noisy, by averaging over a group of neighboring points it is easy
to extract the slope with good accuracy. Note that the graph conforms
with the construction of the initial state described above. This confirms
the validity of the initialization procedure. Note also that on small
scales $(k>10)$ the slope flattens to $0$ showing that, for wavelengths
less than the inter-particle spacing, the power spectra is simply
a white noise as anticipated.

\begin{figure}[ht]
 \centerline{\includegraphics[width=1\textwidth]{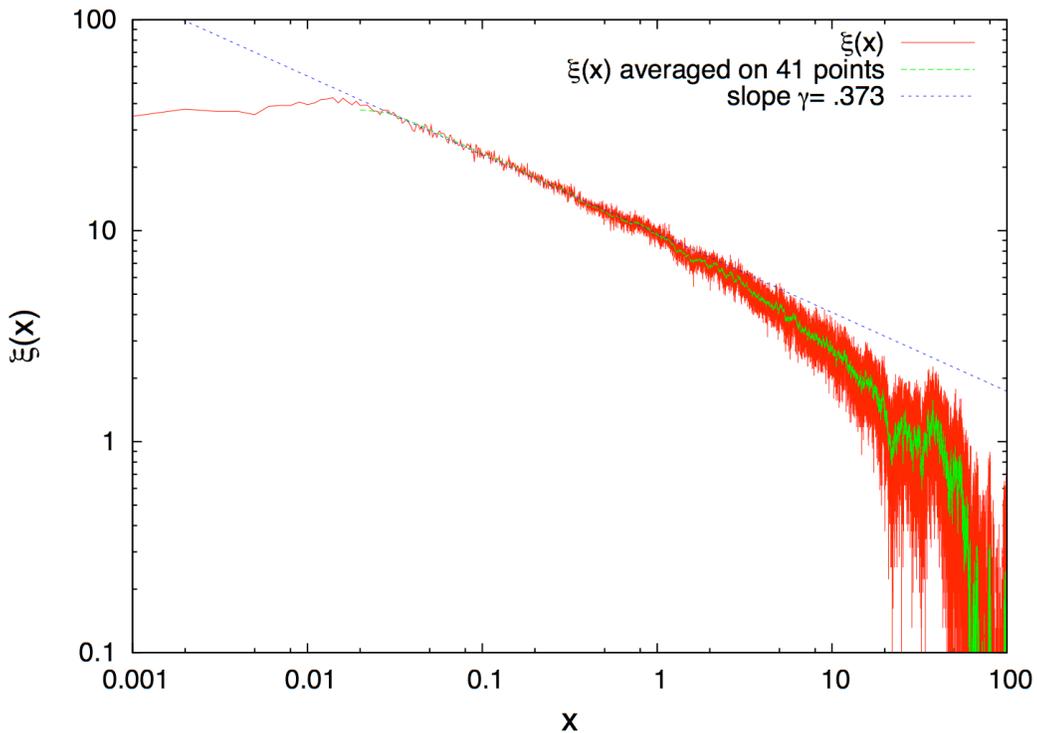}}
\caption{\label{cor} The correlation function at $\tau=13$ for the quintic
model. It exhibits a scaling region in $r$ from about 0.02 to about
2, a range of about 2 decades with a scaling exponent $\gamma=.373$.}

\end{figure}

Historically, power-law behavior in the tail of the density-density
correlation function has been taken as the most important signature
of self-similar behavior of the distribution of galaxy positions \cite{Martinez,cosmo-rev}.
Even at $\tau=18$ we can see from the snapshots that the clusters
are well developed, but still much smaller than the system size. Then
it is safe to assume that boundary conditions do not yet play a significant
role. In figure \ref{cor} we provide a log-log plot of the correlation
function $\xi(r)$ at $\tau=13$ defined by \[
\xi(r)=\left\langle \delta\rho(x+r)\delta\rho(x)\right\rangle \]
 \begin{equation}
\sim\frac{1}{\Delta}\int_{r}^{r+\Delta}\left[\sum_{i,j,i\neq j}\delta(r'-(x_{i}-x_{j}))-\frac{N\left(N-1\right)}{2L^{2}}\right]dr'\label{cor_eq}\end{equation}
 where particles $i$ and $j$ are such that $L/4<x_{i}<3L/4$ to
avoid boundary effects, and $\Delta$ is the bin size. We observe
three distinct regions in figure \ref{cor}. First there is a
relatively flat region at small scales. It is followed by a scaling
region from about $0.02<r<2.0$, a range of about 2 decades, where
$\xi(r)\propto r^{-\gamma}$, which finally deteriorates into noise.
This is similar to cosmological studies which also exhibit the
intra-cluster structure at small scales, and the inter-cluster
scaling seen here. In addition, in the real cosmological setting,
there is a peak due to baryonic acoustic oscillations
\cite{Martinez}. The appearance of noise at larger scales occurs
because the correlation function is decreasing while the noise is
relatively constant. In a log-log plot, as $r$ is increased, at some
point the noise becomes as large as the mean so large fluctuations
appear in the data. Expressed differently, in a log-log plot of the
correlation function, the apparent noise is effectively magnified
with increasing $r$ until it dominates the mean. A straightforward
linearization shows that the effective amplitude of the noise in
$\ln C$ grows as $\exp(\gamma\ln r)$ while it is still relatively
small. In principle it should be possible to reduce the noise by
taking an ensemble average of $\xi(r)$ over many runs, and then
computing the slope of $\ln\left\langle \xi(r)\right\rangle $.
However, since we already employ $2^{16}$ particles, and the
effective noise is growing exponentially with increasing $r$, this
may only slightly extend the apparently noise-free region. By
computing the slope $\gamma$ of the log-log plot in figure \ref{cor}
we are able to obtain an estimate of the correlation dimension
$D_{2}=1-\gamma$ for a one-dimensional system \cite{Bal}. We find
that $\gamma=0.373$ for a scaling region of about two decades in
$l$. This suggests that the correlation dimension is approximately
0.63, which is in agreement within the standard numerical error with
the multifractal analysis described below in some detail (see table
\ref{tab}).

\begin{table}[htdp]
 \caption{\label{tab} Exponent values for the three different models. For pure
power law behavior $D_{2}=1-\gamma=-n$. For each system, $N=65535$
and, initially, $n=2$.}

\begin{indented}

\item[]\begin{tabular}{cccccc}
\br Model  & Time $T$  & $\gamma$  & $n$  & $D_{2}$  & \tabularnewline
\mr Q  & $13$  & .373  & -.672  & .623  & \tabularnewline
 & $16$  & .349  & -.637  & .628  & \tabularnewline
RF  & $13$  & .552  & -.490  & .478  & \tabularnewline
 & $16$  & .547  & -.496  & .500  & \tabularnewline
H  & $13$  & .173  & -1.048  & .862  & \tabularnewline
 & $16$  & .163  & -.880  & .853  & \tabularnewline
\br  &  &  &  &  & \tabularnewline
\end{tabular}\end{indented}
\end{table}

\begin{figure}[ht]
 \centerline{\includegraphics[width=1\textwidth]{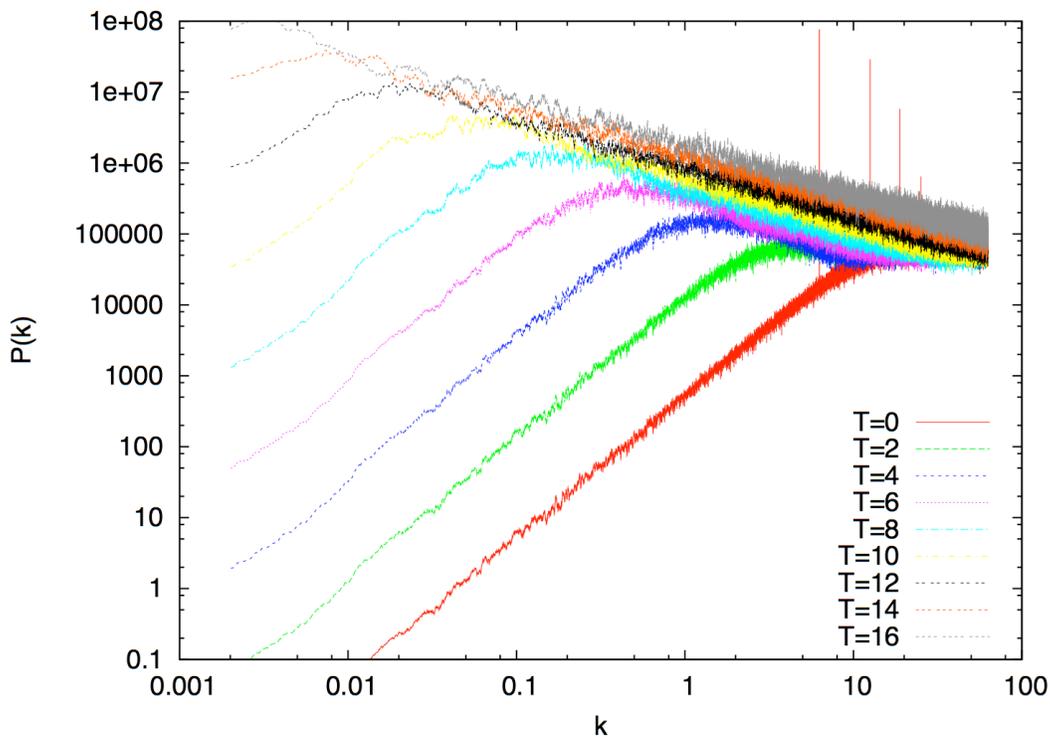}}
\caption{\label{finspec} Plots of the Power spectrum at
$\tau=0,2,4,6,8,10,12,14,16$ for the quintic model. Two scaling
regions are observed separated by the transition from linear to
nonlinear behavior. }

\end{figure}

We have followed the evolution of the power spectra of the density
fluctuations in these systems. In figure \ref{finspec} we display
log-log plots of $P(k)$ for a sequence of times. We see from the
plot that two scaling regimes are present. For small $k$,
\textit{i.e.} on large scales, the system appears to retain the same
dependence on $k$ as in the initial state and the linear regime. In
contrast, using the same averaging method as earlier, we find that a
second power law dependence develops at large $k$. At $\tau=13$ ,
for $k$ less than $.02$ to more than 50, \textit{i.e.} over a range
greater than three decades, this power-law behavior is clearly
exhibited. As time progresses the division between the two regimes
moves to the left so the shape is preserved and the process is
self-similar. Between the two scaling regions is a transition region
centered at, say, $k=k_{c}(\tau)$ which we define below. By
$\tau=16$ there is no longer any trace of the linear regime. The
behavior for $k<k_{c}$ is a remnant of the inter-cluster regions
where the particles have not yet crossed. From Eq. (\ref{eq:linreg})
and the following discussion we see that the initial amplitude of
deviations from the particle equilibrium positions in the growing
mode is increasing as $\exp(2\tau/\sqrt{6})$ resulting in the
corresponding power spectra in the linear region to increase as
$\exp(4\tau/\sqrt{6})$. Analysis of the plots in Fig.
(\ref{finspec}) confirm this nicely. This amplification corresponds
to the {}``stretching'' in the low density regions of $\mu$ space
predicted by Vlasov theory \cite{MRexp,MRGexp}. It is well known,
and discussed from a fluid picture elsewhere in the literature
\cite{peebles2,joyce_shuflat,Joyce_range}.

Notice that in Fig. (\ref{finspec}), as a rough approximation, the
power spectra in the nonlinear region has the form
$P_{nl}(k)=Bk^{n}$ where $B$ is approximately constant and
$n\simeq-.65$. Therefore, in both the linear and nonlinear regimes,
the power spectra are well characterized. We can use this rough
picture to extract useful information about the system evolution.
According to the above, for the linear region we can write
$P_{l}(k)=A\,\exp(\tau/\tau_{l})\, k^{n_{l}}\mbox{ }$where here
$n_{l}=2$, for the quintic model $\tau_{l}=\sqrt{6}$/4, and $A$ is a
constant. Equating $P_{l}(k_{c})=P_{nl}(k_{c})$ at the transition we
find
\begin{equation}
k{}_{c}(\tau)=\left[\frac{B}{A}\right]^{\frac{1}{n_{l}-n}}\exp(-\tau/\tau_{l}(n_{l}-n)).\label{eq:ktran}\end{equation}
In practice, Eq.(\ref{eq:ktran}) describes the transition, defined
as the intersection of the two power laws, remarkably well. Within
the context of this model, $2\pi/k_{c}$ represents the size of the
largest clusters at that epoch. If we further assume that the
intra-cluster particle density remains roughly constant, then we
also see that the number of clusters deceases approximately as
$\exp(-\tau/\tau_{m})$ where $\tau_{m}=\tau_{l}(n_{l}-n)$ can be
taken as a measure of the mean time between mergers.

We can also obtain insight into the fractal nature of the distribution
of particles in configuration space from this simple model. From the
Weiner-Khinchine Theorem \cite{Gard}, the correlation function $\xi(r)$
can be obtained from the Fourier transform of P(k):
\begin{equation}
\xi(r)=\frac{1}{2\pi}\int dk\, P(k)\, cos(kr).\label{eq:wk}
\end{equation}
Combining the contributions on either side of $k_{c}$ we obtain after
a little algebra
\begin{equation}
\xi(r)=\frac{1}{2\pi}Bk_{c}^{1+n}\left\{ \int_{0}^{1}du\, u^{2}cos(rk_{c}u)+\int_{1}^{\infty}du\, u^{n}\, cos(rk_{c}u)\right\} .\label{eq:cormod}
\end{equation}
The correlation dimension is arguably the most robust of the class
of generalized fractal dimensions (see below). We can obtain the correlation
dimension by examining the average distribution of particles $C(R)$
in a ball of radius $R$ around a given particle \cite{Ott,Bal},\begin{eqnarray}
C(R) & = & \rho\int_{-R}^{R}dr(1+\xi(r))\end{eqnarray}
\begin{equation}
=2\rho\left\{ R+\frac{1}{2\pi}Bk_{c}^{n}\left[\int_{0}^{1}du\, u\, sin(Rk_{c}u)+\int_{1}^{\infty}du\, u^{n-1}\, sin(Rk_{c}u)\right]\right\} .
\end{equation}
The correlation dimension is given by
\begin{equation}
D_{2}=\lim_{R\rightarrow0}\frac{\ln C}{\ln R}=\lim_{R\rightarrow0}\frac{1}{\ln R}\ln\left[R+\frac{B}{2\pi}\left(a_{l}k_{c}^{n+1}R+a_{nl}R^{-n}\right)\right]
\end{equation}
where, in the above, $\rho$ is the average density, the second
integral was expressed in terms of the incomplete Gamma function,
only the lowest order contributions to $C(R)$ were included and
$a_{l}$ and $a_{nl}$ are constants of order unity. We see that when
$\tau\ll\tau_{m}/(n+1)$, {\it i.e.} early in the evolution, the
linear term dominates and we anticipate that numerical analysis of
the simulations will yield $D_{2}\simeq1$ while, for late times, the
last term, contributed by the nonlinear regime, is dominant and
$D_{2}\simeq-n.$

Careful observation of the clusters in the phase plane suggests a
possible mechanism for the hierarchical, bottom-up cluster formation.
The large amplitude on small scales initiates a large collection of
small clusters. Perhaps they are seeded on the caustics studied by
Yano and Gouda\cite{Gouda_caus}. Once these clusters virialize, they
{}``freeze out'' of the expansion and become the new {}``particles''
for the next epoch of larger clusters \cite{peebles2}. In turn, these
cluster {}``particles'' virialize and merge and the process continually
repeats until boundary effects interfere. At large scales $P(k)$
continues to increase as a result of the exponential growth in the
remnants of the growing mode. On the other hand, at smaller scales,
the freezing out of the clusters blocks the steady increase of $P(k)$.
Thus virialization on successively larger scales determines the shape
of the power spectra and qualitatively explains the observed results
of the simulations. Since it is a renormalization process, the evolution
is self-similar. Of course, it is a continuous process. The fact that
larger {}``sub-clusters'' become the particles for the next stage
of virialization causes the clusters to grow in real space. In their
study, Yano and Gouda \cite{Gouda_powsp} found three scaling ranges,
the linear regime for large scales, a small intermediate range with
an index of $-1$, and the third region which they deemed multi-caustic.
They hypothesize that the power-law behavior is induced by the generation
of caustics following the first crossings. The small regime, with
$n=-1$, is associated with the period of single crossings. In our
simulations we see hints of the intermediate regime in the transition
regions of Fig. (\ref{finspec}). In addition the index values we
obtained for the behavior on small scales approximately correspond
to theirs. Since Gabrielli \etal have argued that the Q model and
the Zeldovich approximation are equivalent \cite{Joyce_1d}, this
isn't surprising. Additional discussion of the transition from the
linear to nonlinear regime can be found in \cite{joyce_shuflat}.

\section{Fractal Measures}

\label{fractal-measure}

It is natural to assume that the apparently self-similar structure
that develops in the phase plane (see figure \ref{snap}) as time
evolves develops fractal geometry, but we will see that things are
not so simple. In their earlier study of the RF model, Rouet, Feix
and Jamin found a box counting dimension for the particle positions
in $\mu$ space of about 0.75 for an initial water bag distribution
(uniform on a rectangle in the phase plane) and a fractal dimension
of about 0.57 in the configuration space (\textit{i.e.}, of the projection
of the set of points in $\mu$ space on the position axis) \cite{Rouet2}.
As far as we know, Balian and Schaeffer were the first to suggest
that the distribution of galaxy positions is consistent with a bifractal
geometry \cite{Bal}. Their idea was that the geometry of the galaxy
distribution was different in the clusters and voids and, as a first
approximation, this could be represented as a superposition of two
independent fractals. Of course, their analysis was restricted solely
to galaxy positions. Since the structures which evolve are strongly
inhomogeneous, here we perform a multifractal analysis \cite{Hal}
of the configuration space.

The multifractal formalism shares a number of features with thermodynamics
\cite{Hal,Ott}. To implement it we partitioned configuration space
into cells of length $l$. At each time of observation in the simulation,
a measure $\mu_{i}=N_{i}(t)/N$ was assigned to cell $i$, where $N_{i}(t)$
is the population of cell $i$ at time $t$ and $N$ is the total
number of particles in the simulation. The generalized dimension of
order $q$ is defined by \cite{Hal}

\begin{equation}
D_{q}=\frac{1}{q-1}\lim_{l\rightarrow0}\frac{\ln{C_{q}}}{\ln{l}},\quad C_{q}=\Sigma\mu_{i}^{q},\label{Dq}\end{equation}
 where $C_{q}(l)$ is the effective partition function \cite{Ott},
$D_{0}$ is the box counting dimension, $D_{1}$, obtained by taking
the limit $q\rightarrow1$, is the information dimension, and $D_{2}$
is the correlation dimension \cite{Hal,Ott}. As $q$ increases above
$0$, the $D_{q}$ provide information on the geometry of cells with
higher population, \textit{i.e.} the regions of high density or clusters.

In practice, it is not possible to take the limit $l\rightarrow0$
with a finite sample. Instead, one looks for a scaling relation over
a substantial range of $\ln l$ with the expectation that a linear
relation between $\ln C_{q}$ and $\ln l$ occurs, suggesting power-law
dependence of $C_{q}$ on $l$. Then, in the most favorable case,
the slope of the linear region should provide the correct power and,
after dividing by $q-1$, the generalized dimension $D_{q}$. As a
rule, or guide, if scaling can be found either from observation or
computation over three decades of $l$, then we typically infer that
there is good evidence of fractal structure \cite{Mac}. Also of interest
is the global scaling index $\tau_{q}$, where $C_{q}\sim l^{\tau_{q}}$
for small $l$. From Eq. (\ref{Dq} we see that $\tau_{q}$ and $f_{(}\alpha)$
are related to each other by $D_{q}(q-1)=\tau_{q}$ for $q\neq1$
\cite{Ott}. Here we present the results of our fractal analysis of
the particle positions on the line (position only).

If it exists, a scaling range of $l$ is defined as the interval on
which plots of $\ln C_{q}$ versus $\ln l$ are linear. Of course,
for the special case of $q=1$, we plot $-\Sigma\mu_{i}\ln\mu_{i}$
\textit{vs} $\ln l$ to obtain the information dimension \cite{Ott}.
If a scaling range can be found, $D_{q}$ is obtained by taking the
appropriate derivative. It is well established by proof and example
that, for a normal, homogeneous fractal, all of the generalized dimensions
are equal, while for an inhomogeneous fractal, \textit{e.g.}, the
Henon attractor, $D_{q+1}\leq D_{q}$ \cite{Hal}. In the limit of
small $l$, the partition function $C_{q}(l)$ can also be decomposed
into a sum of contributions from regions of the inhomogeneous fractal
sharing the same pointwise dimension $\alpha$,

\begin{equation}
C_{q}(l)=\int d\alpha\, l^{\alpha q}\rho(\alpha)\, l^{-f(\alpha)}\label{cq}\end{equation}
where $f\left(\alpha\right)$ is the fractal dimension of its support
\cite{Hal,Fed,Ott}. Then if, for a range of $q$, a single region
is dominant, we find a simple relation between the global index $\tau_{q}$
and $\alpha$,

\begin{equation}
\tau_{q}=\alpha q-f(\alpha).\label{tauq}\end{equation}
 and a corresponding linear relation between $\ln C_{q}$ and $q$.

Initially the system is very cold, the $\mu$ space distribution of
particles appears as a line, and the initial dimension is also about
unity. In fact, initially the velocity of the particles is a perturbation.
It is not large enough to allow particles to cross the entire system
in a unit of time. After a while the fluctuations of force in the
system destroys the approximate symmetry of the initial $\mu$ space
distribution. Breaking the symmetry leads to the short-time dissipative
mixing that results in the separation of the system into clusters.
The behavior of the distribution of points in configuration space
is similar. The initial dimension is nearly unity until clustering
commences. At this time the dimensions in $\mu$ space and configuration
space separate.

As time progresses, however, for the initial conditions discussed
above, typically one dominant scaling region and a hint of a second
develops. Of course this is in addition to the trivial scaling regions
obtained for very small $l$, corresponding to isolated points, and
to large $l$ on the order of the system size, for which the matter
distribution looks smooth. The observed size of each scaling range
depended on both the elapsed time into the simulation and the value
of $q$. In some instances it was possible to find good scaling up
to $4.7$ decades in $l$ !

\begin{figure}[ht]
 \centerline{ \subfloat{(a)}{\includegraphics[width=0.5\textwidth]{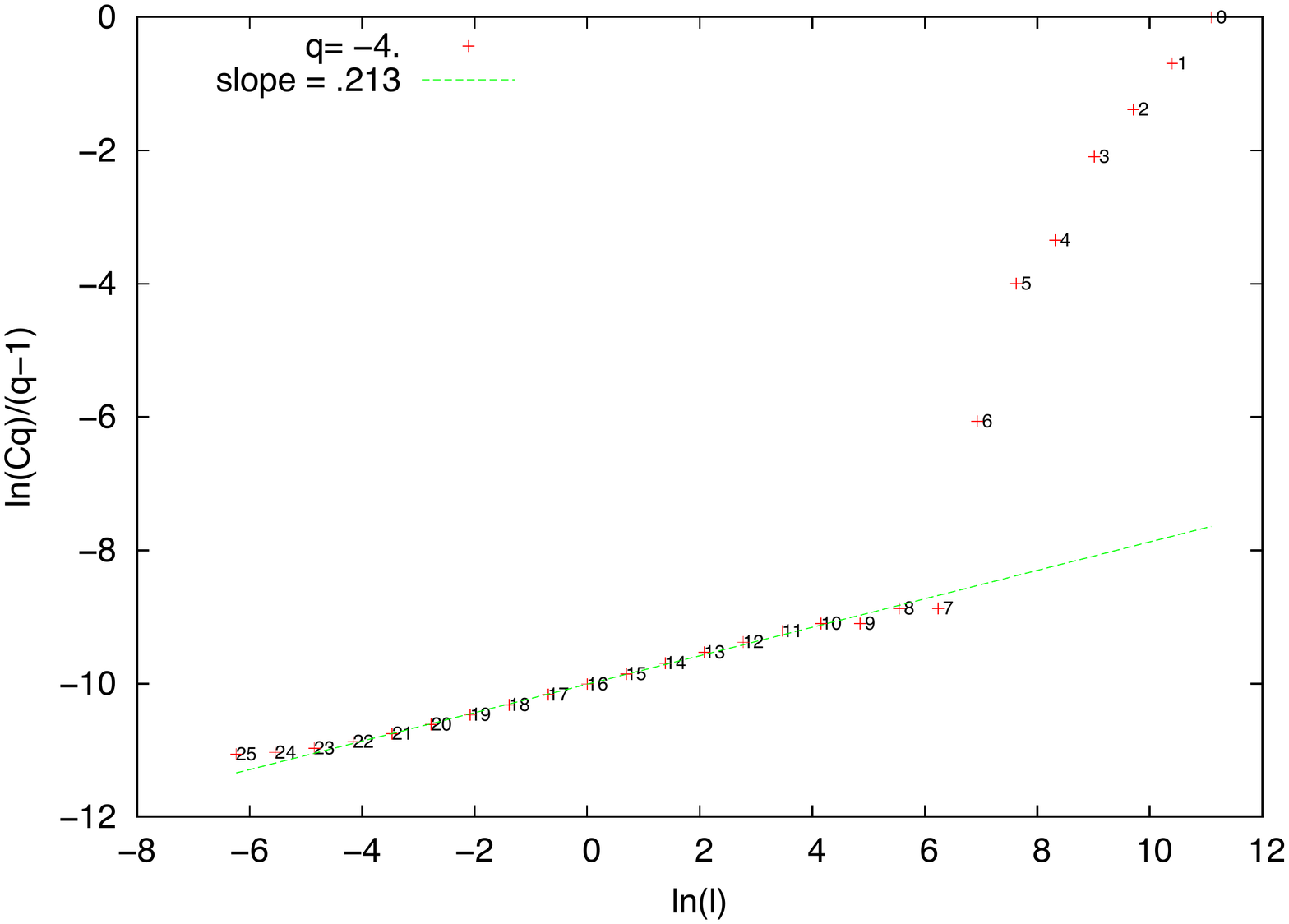}}
\subfloat{(b)}{\includegraphics[width=0.5\textwidth]{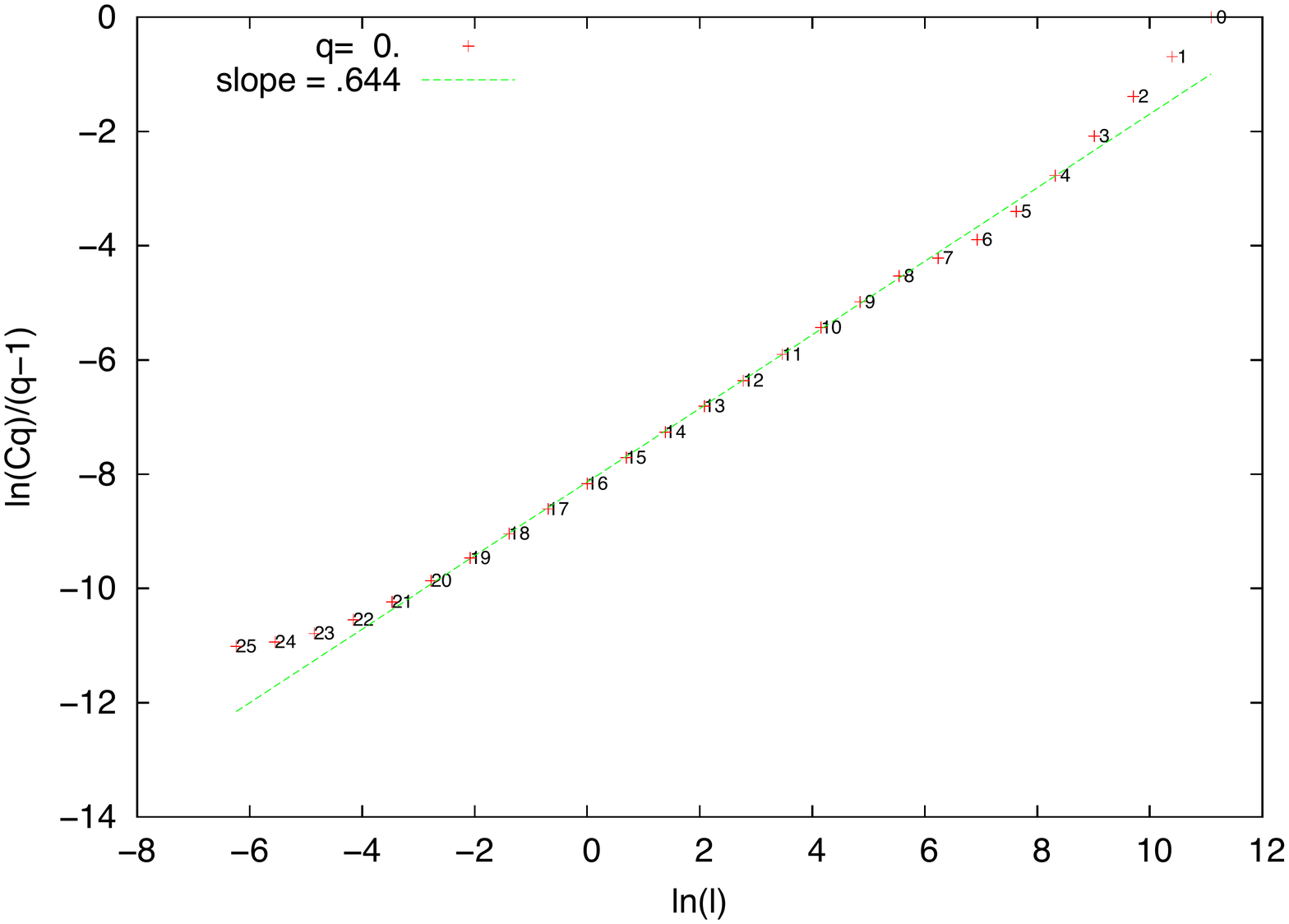}}}
\centerline{ \subfloat{(c)}{\includegraphics[width=0.5\textwidth]{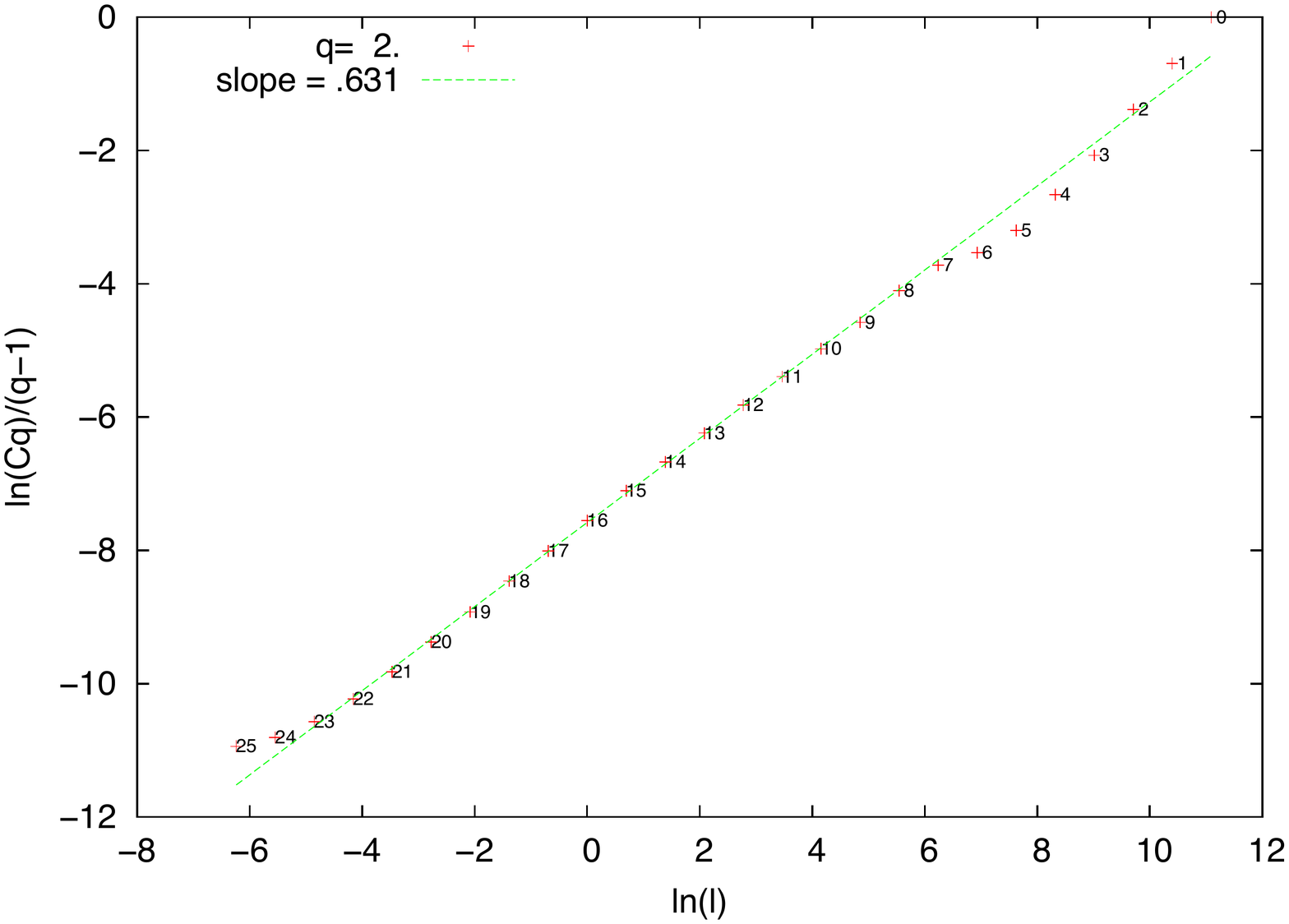}}
\subfloat{(d)}{\includegraphics[width=0.5\textwidth]{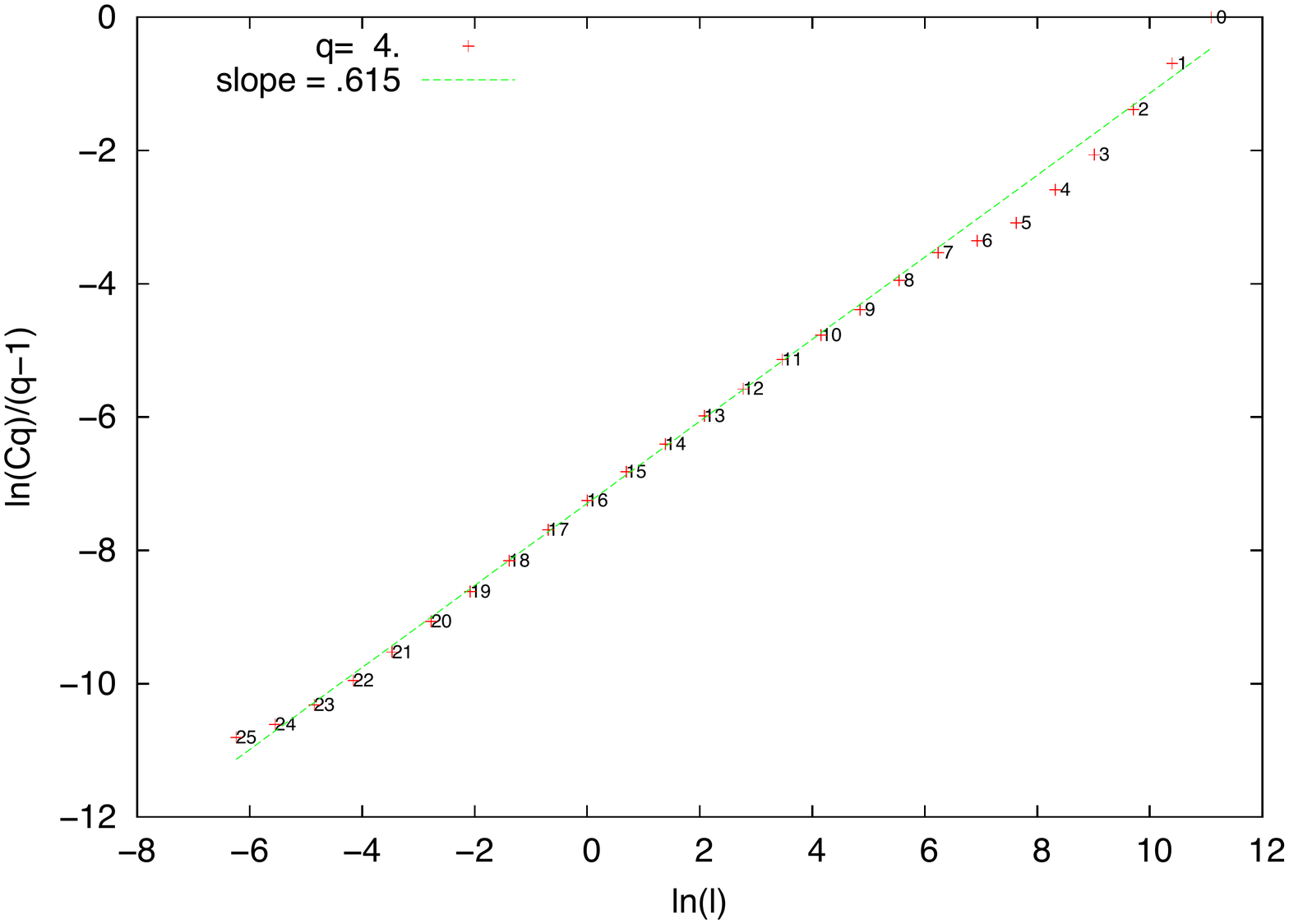}}}
\caption{\label{scal} Scaling behavior in configuration space at $\tau=16$
for the quintic model. Plots of $\frac{1}{q-1}{\ln{C_{q}}}$ versus
$\ln{l}$ are provided for four values of q: a) q=-4, b) q=0. c) q=2,
d) q= 4. The slopes have been determined using a linear regression
from points 12 to 20.}

\end{figure}

In figure \ref{scal} we provide plots of $\frac{1}{q-1}\ln C_{q}$
versus $\ln l$ in configuration space for four different values of
$q$ covering the range we investigated, ($-4<q<4$). To guarantee
that the fractal structure was fully developed, we chose $\tau=16$
for the time of observation and the initial conditions are those given
above. There was a basic consistency in the results with the largest
dominant scale $-5.5<\ln l<6.5$ occurring for $q=4$ and the smallest,
$-4<\ln l<4$, for $q=-4,$ \textit{i.e.} from about $3.5$ to 5 decades.
In all cases there was a hint of a second scaling region just beyond
the primary, \textit{e.g.} for $6.2<\ln l<7.6$ for the case $q=4.$
This scale corresponds to the turnover point in the power spectra.
It appears to be a manifestation of the remnant of the linear region
at large scales. The straight (dashed) lines that appear in the data
are best curve fits obtained from linear regression for the range
of the larger scaling region that is common to each of the values
of $q$. Their slopes were used to compute the generalized dimensions.

\begin{figure}[ht]
 \centerline{\includegraphics[width=1\textwidth]{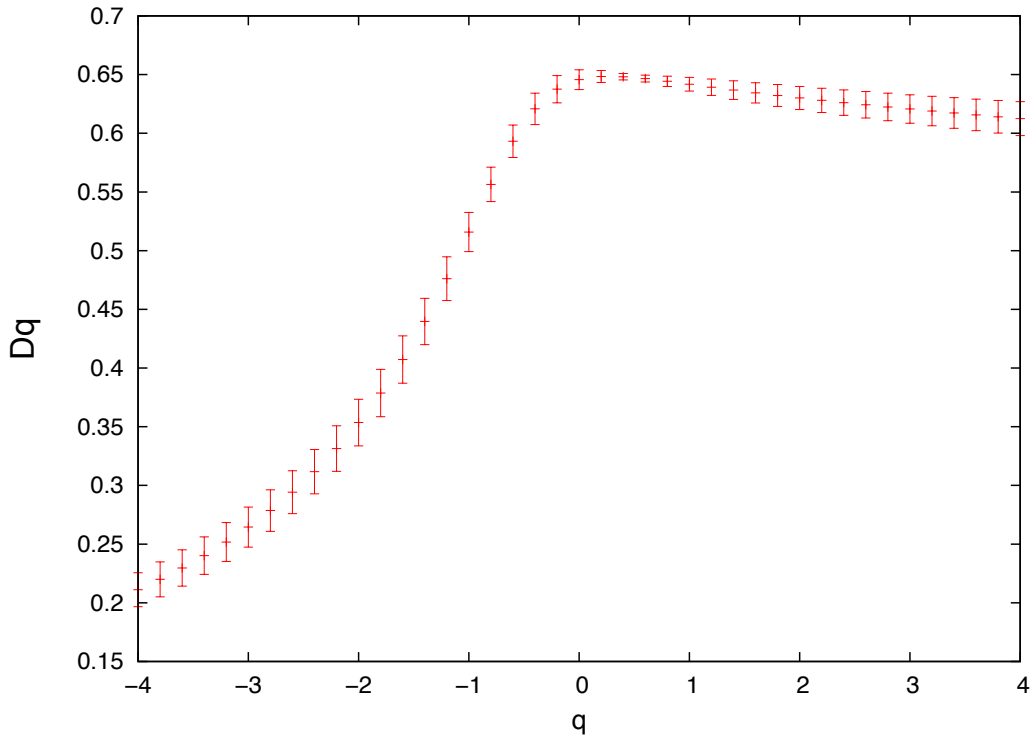}}
\caption{\label{dq} Generalized dimension $D_{q}$ \textit{vs} $q$ in configuration
space for the quintic model at $\tau=16$.}

\end{figure}

\begin{figure}[ht]
 \centerline{\includegraphics[width=1\textwidth]{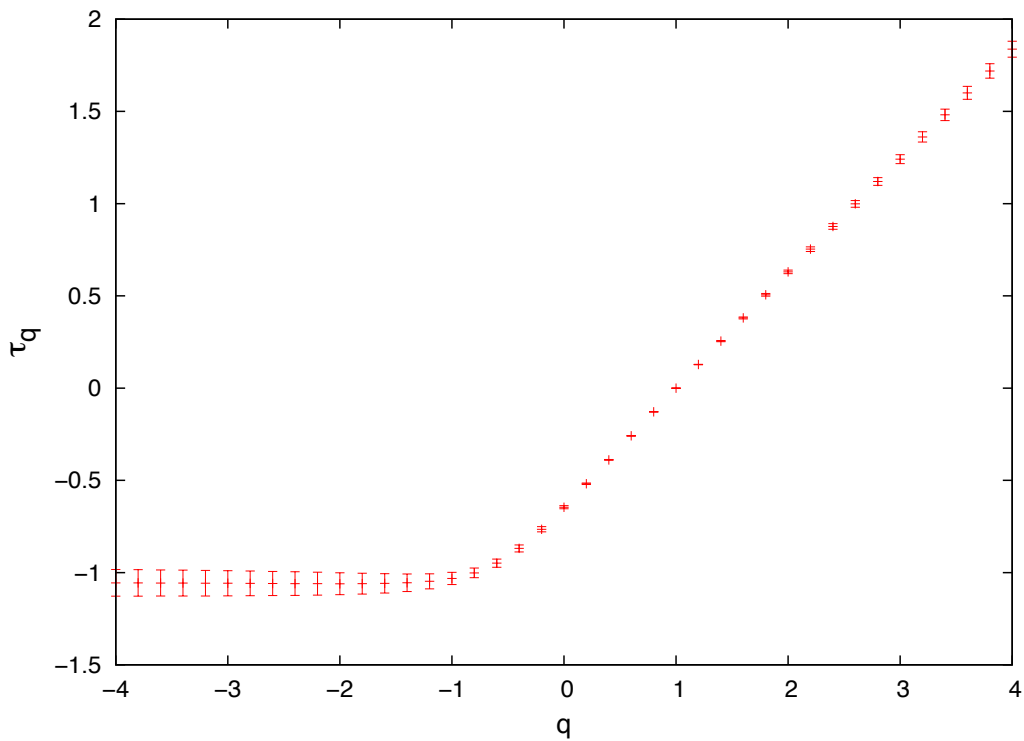}}
\caption{\label{tau_q} Global scaling index $\tau_{q}$ \textit{vs.} $q$
in configuration space for the quintic model at $\tau=16$.}

\end{figure}

In figure \ref{dq} we illustrate the behavior of the generalized
dimension $D_{q}$ for the configuration space of the quintic model
at $\tau=16$ for $-4<q<4$. Although the embedding dimension is $d=1$,
$D_{0}$ is about $0.65$ so the distribution is definitely fractal.
For $q<0$ , the curve is an increasing function. This unphysical
behavior suggests that there is simply a lack of sufficient data in
the low-density regions to provide reliable dimensions. On the other
hand, for $q>0$ , we see the decreasing behavior characteristic of
a multifractal. In figure \ref{tau_q} we present a plot of $\tau_{q}$
\textit{vs} $q$ constructed from the same data. Note the constant
value for $q<0$. Comparison with equation (\ref{tauq}) suggests
that the pointwise dimension vanishes in the under-dense regions and
that there is simply insufficient data to fix the type of geometry
\cite{Ott}. This is consistent with figure \ref{dq}. For the case
where $q>0$, the nearly linear behavior suggests that equation (\ref{tauq})
applies here as well and yields a pointwise dimension $\alpha$ of
about 0.65, and a support with Hausdorff dimension $f(\alpha)$ of
about 0.7. It is not immediately obvious how to reconcile figures
\ref{dq} and \ref{tau_q} for $q>0$ but it is interesting that both
$\alpha$ and $f(\alpha)$ are very close to the value of $D_{0}$.
In varying the initial conditions by selecting different random number
seeds, there are slight variations in the overall picture. For $q>0$
we find $D_{q}$ consistently in the neighborhood $[0.62,\:0.65]$
but the degree of the multi-fractal nature, gauged by the negative
slope in figure \ref{dq}, varies somewhat. Perhaps the multifractal
character suggested by the decrease of $D(q)$ with increasing $q$
is an artefact due to the increasing sparseness of denser and denser
regions. However, since simulations of the RF model exhibit similar
structure formation and a flat $D(q)$, this is doubtful.

Part of our goal is to compare how fractal geometry arises in a family
of related models. So far we have presented results for the quintic,
or Q, model. However we have also carried out similar studies of the
Rouet-Feix (RF) model and the Hamiltonian model without friction (which
can be obtained from either of the former by nullifying the first
derivative contribution in equation (\ref{eq:final})). In table \ref{tab}
we compare the correlation dimension $D_{2}$ and the exponents generated
by the power spectrum and correlation function for all three models
at scaled times $\tau=13$ and $16$, where the clustering is highly
developed. Writing $P(k)\sim k^{n}$ and $\xi(r)\sim r^{-\gamma}$,
in principle there are simple relations between $D_{2}$, $n$, and
$\gamma$ ; namely $D_{2}=1-\gamma=-n$. The first equality can be
easily derived from the definition of correlation dimension (see \textit{e.g.}
\cite{Bal,Ott}and the discussion in the previous section), while
the second is due to the Wiener-Khintchine theorem, \textit{i.e.}
the power spectrum is the Fourier transform of the correlation function
\cite{Gard}. We see from table \ref{tab} that these relations are
approximately obeyed in these systems with $2^{16}$ particles. The
agreement is superior for the Q and RF models. Experience has shown
that, for them, as the system size is increased, so also is the agreement.
While there are similarities in the fractal structure of the RF and
Q models, they are not the same. For example in the quintic model
the generalized fractal dimension $D_{q}$ is consistently greater
than the corresponding dimension in the RF model showing that the
inhomogeneity is stronger in the latter although the larger friction
constant of the RF model causes clustering to commence later in the
system evolution. Another difference is that plots of $D_{q}\: vs\: q$
for positive $q$ for the RF model are virtually constant, suggesting
monofractal behavior. The values obtained for $n$ are consistent
with those found in the simulations carried out by Yano and Gouda
using the Zeldovich approximation in the {}``multicaustic'' regime
\cite{Gouda_powsp}.

\begin{figure}[ht]
 \centerline{\subfloat{(a)}{\includegraphics[width=0.5\textwidth]{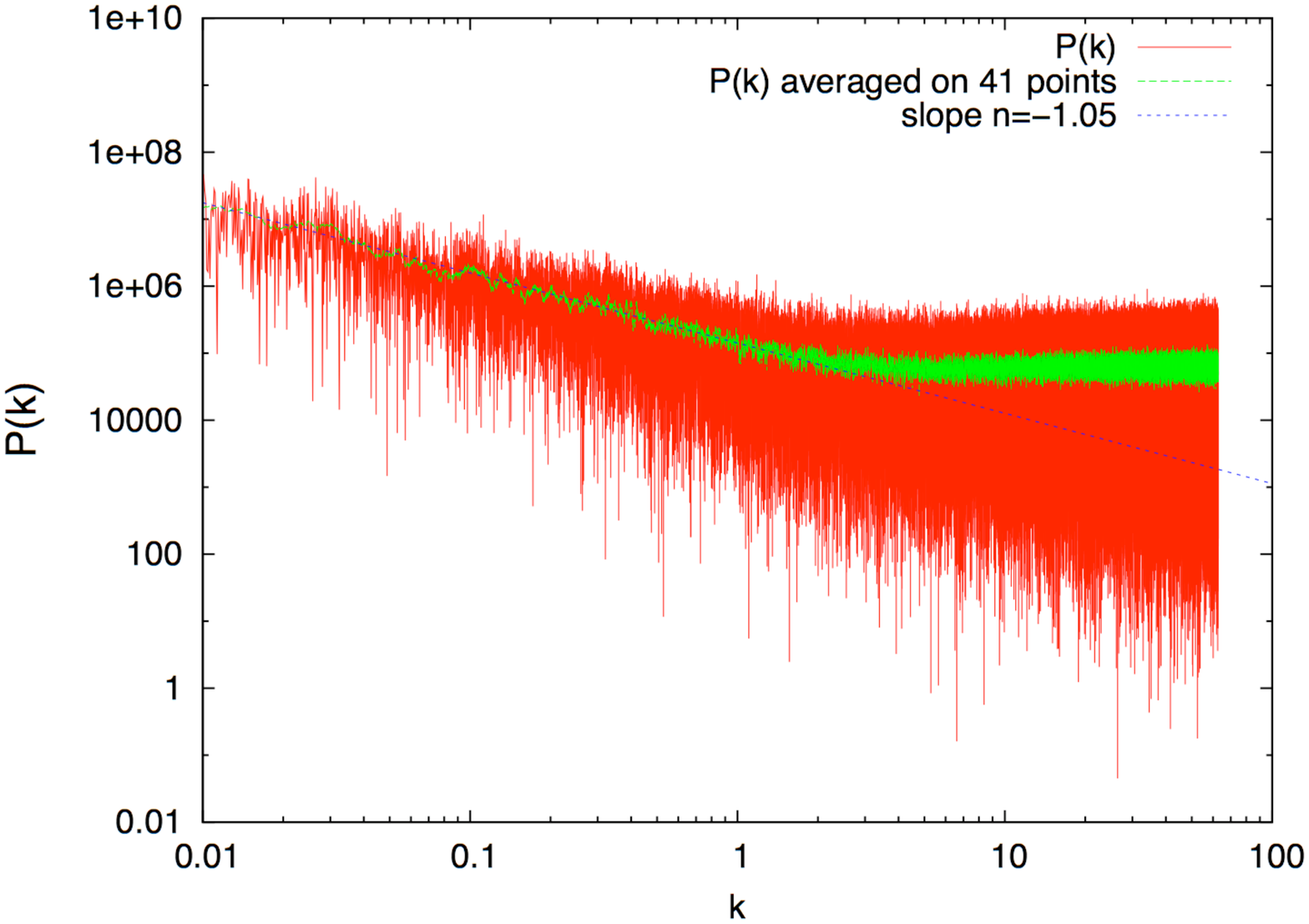}}
\subfloat{(b)}{\includegraphics[width=0.5\textwidth]{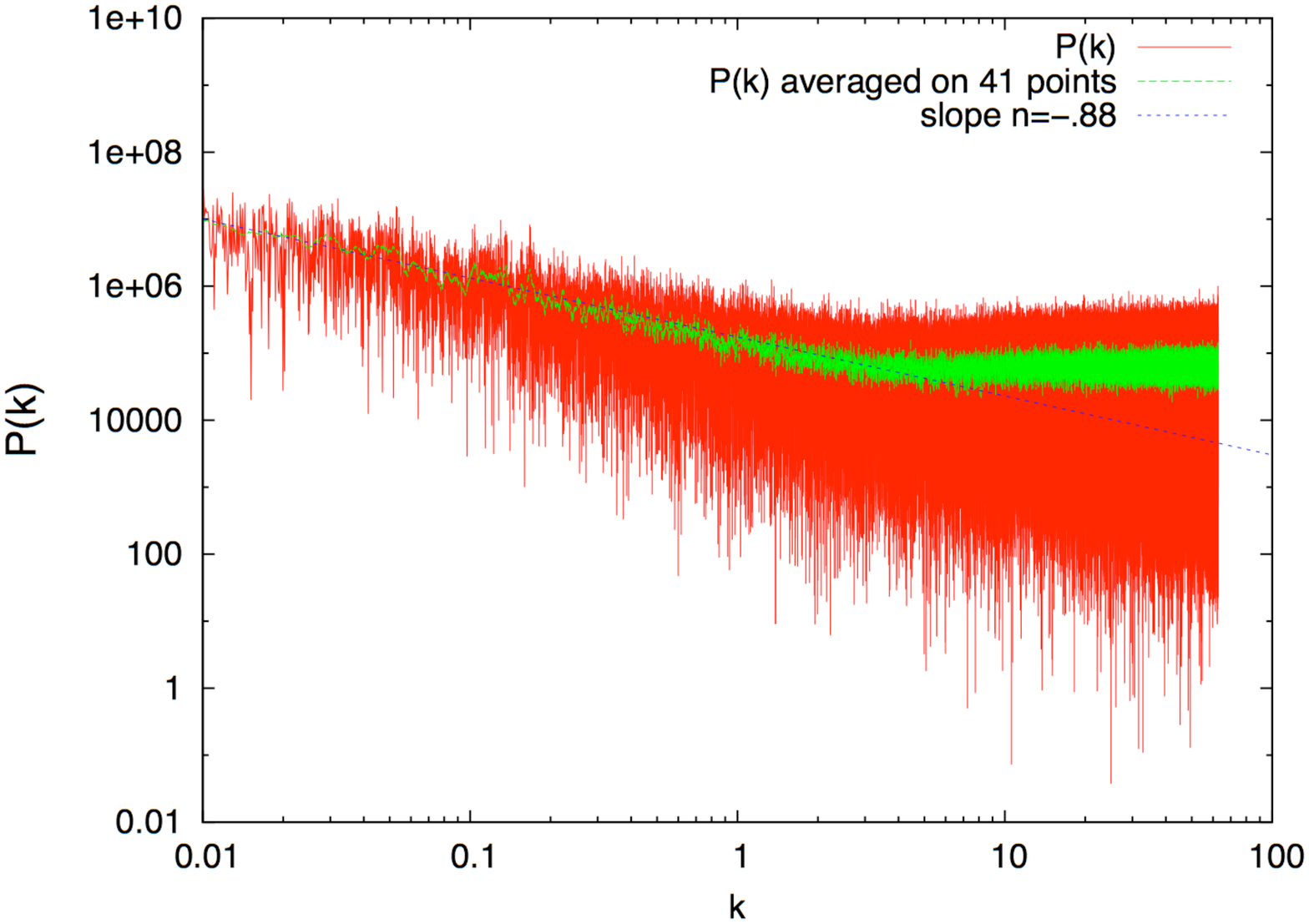}}}
\caption{\label{Hamps}The Power spectrum for the Hamiltonian model : (a) at
$\tau=13$ it exhibits a scaling region in $k$ from about 0.01 to
1, with a scaling exponent $-n=1.05$ (b) at $\tau=16$, the scaling
range is {[}0.01;2{]}, and $-n=0.88$.}

\end{figure}

\begin{figure}[ht]
 \centerline{\includegraphics[width=1\textwidth]{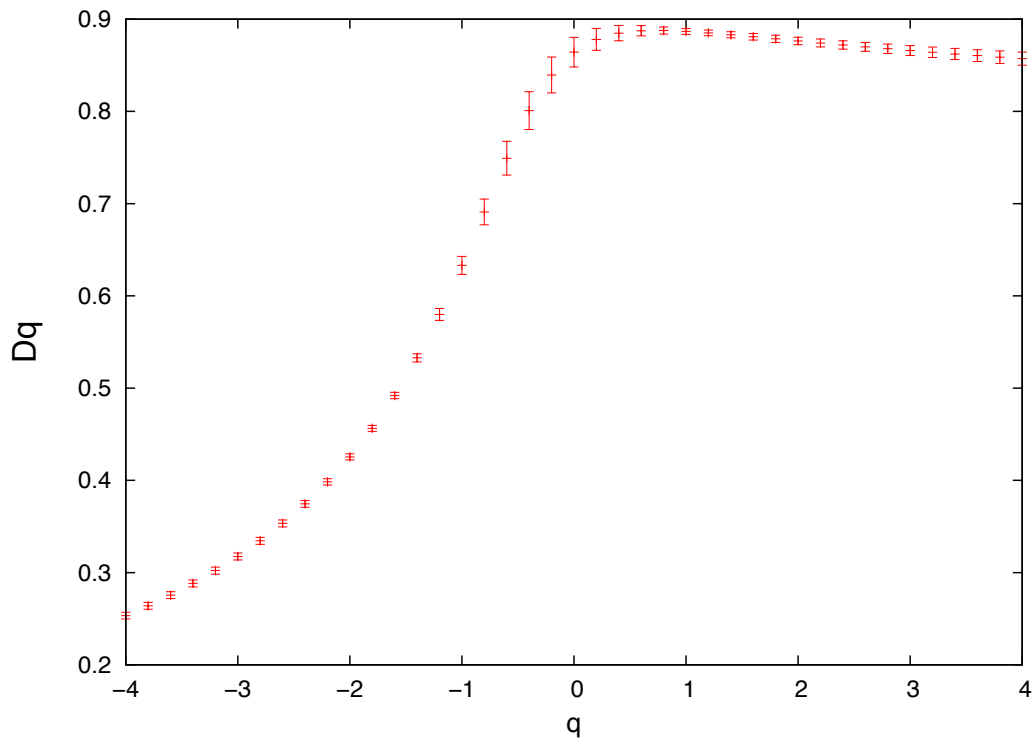}}
\caption{\label{DqH} Generalized dimension $D_{q}$ \textit{vs} $q$ in configuration
space for the Hamiltonian model at $\tau=16$.}

\end{figure}

\begin{figure}[ht]
 \centerline{\includegraphics[width=1\textwidth]{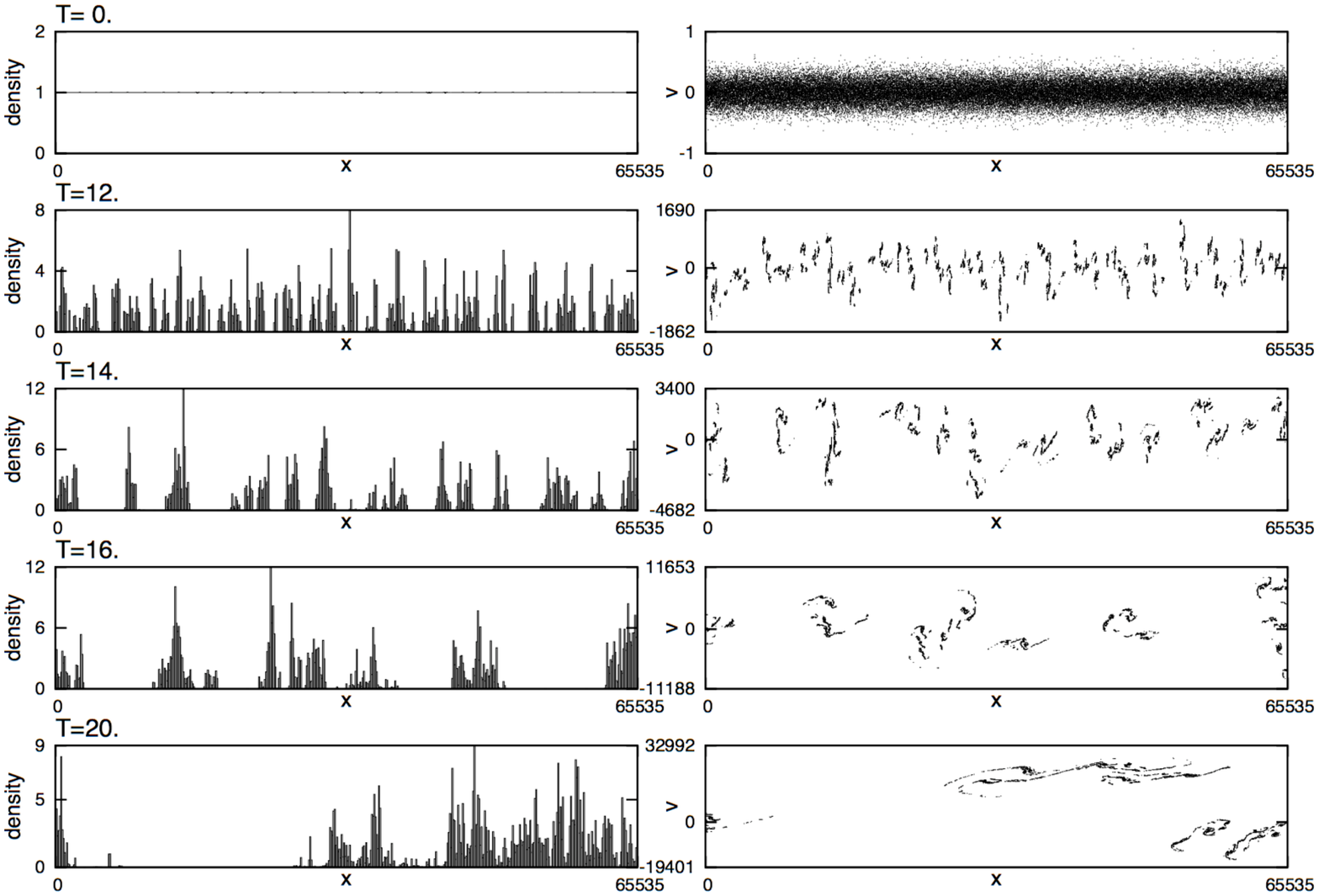}}
\caption{\label{Hsnap} Evolution in configuration and $\mu$ space for the
Hamiltonian model with $2^{16}$ particles from $\tau=0$ to $\tau=20$.
The initial distribution is such that the density power spectrum has
a power-law of index $n=2$.}

\end{figure}

The Hamiltonian system was also investigated by Gabrielli \etal \cite{Joyce_1d}.
Of the three models, it exhibits the most rapid cluster formation.
However, in other aspects the behavior differs sharply from the dissipative
systems. First, the scaling behavior is less robust. As shown in figures
\ref{Hamps}, plots of the power spectra at $T=13,\:16$ have a smaller
scaling range. In addition, the fractal properties are much closer
to an ordinary system. In figure \ref{DqH} we see that the generalized
dimensions $D_{q}\simeq0.9$, much greater than the RF model, which
presents the greatest structure with $D_{q}\simeq0.5$. As can be
seen in table \ref{tab}, the connection between the exponents is
weaker, no doubt due to the smaller scaling range. The relation between
the exponents is only exact if the correlation function and power
spectra are power-laws. Finally, if we extend the simulations beyond
the time where boundary conditions can be neglected, say for $T>16$,
we see a remarkable difference. For such times the matter distribution
in the dissipative systems eventually collapses to just a single cluster.
In contrast, as we see in figure \ref{Hsnap}, the Hamiltonian system
becomes space filling. It appears to revert to the stationary state
for this system introduced by Valageas \cite{Val1,Val2}.

\begin{figure}[ht]
 \centerline{\includegraphics[width=1\textwidth]{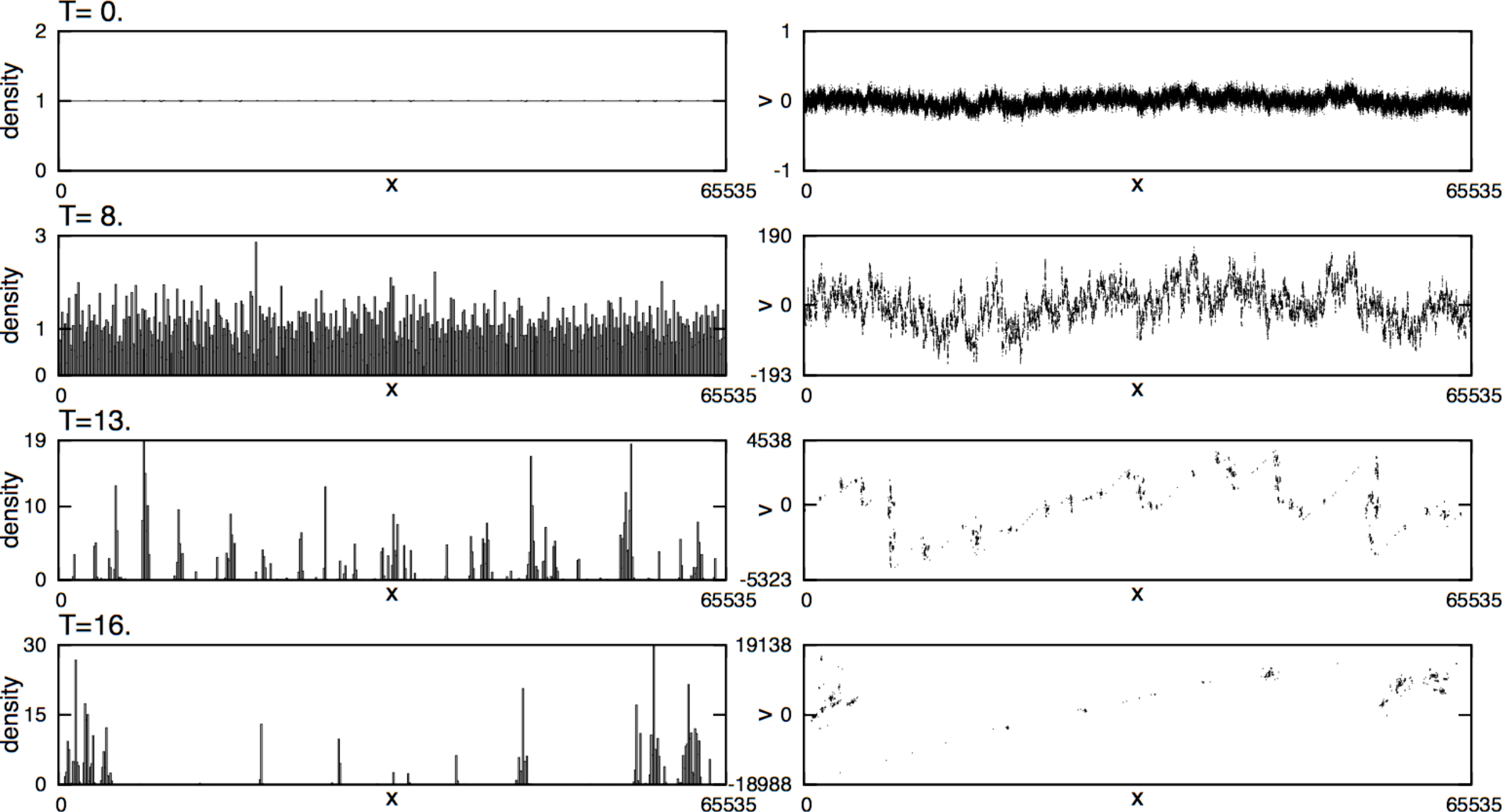}}
\caption{\label{n=00003D00003D1} Evolution in configuration and $\mu$ space
for the quintic model with $2^{16}$ particles from $\tau=0$ to $\tau=16$.
The initial distribution is such that the density power spectrum has
a power-law of index $n=1$.}

\end{figure}

\begin{figure}[ht]
 \centerline{\includegraphics[width=1\textwidth]{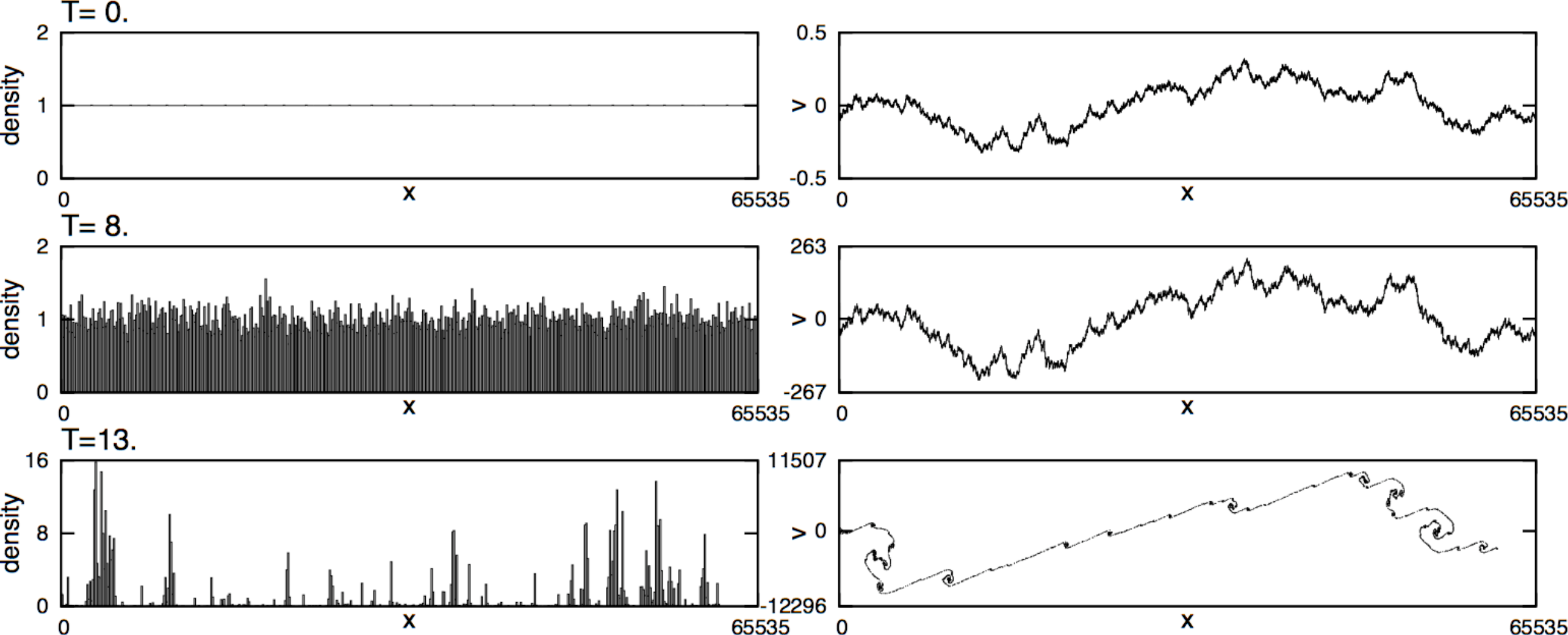}}
\caption{\label{n=00003D00003D0} Evolution in configuration and $\mu$ space
for the quintic model with $2^{16}$ particles from $\tau=0$ to $\tau=13$.
The initial distribution is such that the density power spectrum has
a power-law of index $n=0$.}

\end{figure}

There is a sensitive dependence of the evolution on the initial conditions
and, in particular, on the initial spectral index $n$. Although the
cluster distributions in configuration space are similar, more information
is available in the phase plane. We find that the smaller the initial
index, the greater is the predominance of large clusters in the initial
state, and these are very unstable. In figures \ref{n=00003D00003D1}
and \ref{n=00003D00003D0} we show snapshots of the evolution of the
Q model with $n=1$ and 0. We see from the figures that there is a
strong correlation between the initial condition and the type of structure
formation that develops. For the case where the initial particle distribution
is white noise ($n=0$), examine the $\mu$ space snapshot at the
time $\tau=13.$ Note that the system has not broken apart as we would
expect from the simulations with $n=2.$ Moreover, the characteristic
bottom-up behavior, where the smaller clusters combine to form larger
ones as time evolves, is missing. Instead there is a single, large,
connected, structure with a few small {}``subclusters'' distributed
within the pattern. Clearly hierarchical clustering is absent. Examining
the snapshots for $n=1$ we see that the behavior is intermediate.
Both small and large clusters are present, but the bottom-up scenario
is not nearly as robust as for $n=2$. A clue for the origin of this
behavior may be found in the work of Gabrielli \etal mentioned earlier
\cite{Gab_Joy_1DPD,Joyce_range}. In their investigation of the statistical
properties of static distributions of infinite systems that are on
average uniform, they found that the variance of the force diverges
if the structure factor is characterized by white noise ($n=0)$ and
is marginal for $n=1$. Alternatively, the source of the behavior
may arise from dynamical considerations. Energy at small scales is
required to establish the presence of small clusters that can initiate
a process of hierarchical virialization. For $n=0$ there is simply
too much competition from large clusters and the hierarchical process
is broken. Although the hierarchical process appears to start in a
few locations, it doesn't have sufficient time to develop. Long wavelengths
drive the overall simulation.

\section{Summary and Conclusions}

\label{summary}

Here and in our earlier work we have seen that hierarchical clustering
is an intricate, highly choreographed dance in phase space that requires
an initially cold and uniform system. The three models that were studied
here differ only in the value of the friction constant, $\alpha$:
$\alpha=$$\frac{1}{\surd2,},$$\frac{1}{\surd6,},0$ , respectively,
for the RF, Q, and Hamiltonian models. The Q (Zeldovich) model has
a hybrid geometry that may actually represent the local behavior of
a 3+1 dimensional system for a short time, but cannot reproduce its
real clustering properties in the highly nonlinear regime. Alternatively,
the RF model is completely self-consistent with the transformation
to comoving coordinates in 1+1 dimensions. It is useful for unambiguously
studying nonlinear behavior, but has a weaker connection to the 3+1
dimensional universe. In fact, simulations of both the Q and RF models
show remarkably similar behavior. The Hamiltonian model provides a
contrasting dynamical system with no dissipation in the comoving coordinate
system. Like the Q and RF models, it undergoes rapid hierarchical
clustering for a short time span. However, unlike them, on long time
scales the system becomes space filling and appears to revert to the
stable stationary state investigated by Valageas \cite{Val1,Val2}.

To mimic cosmological investigations, the initial positions were
selected by sampling a power-law spectrum with random phases. The
velocities were chosen to insure that the system is in the growing
mode in phase space. The three models were studied with a range of
initial power-law indices $n$. Qualitative behavior was observed for
$n=3,4$ while extensive studies were carried out for $n=0,\:1,\:2$.
Both the RF and Q models exhibited very similar behavior. For the
case $n=2$, initially the systems remain fairly uniform. After some
time, around $\tau=8$ for the Q model, the systems break into small
clusters. As time progresses further the clusters coalesce into
larger ones, supporting the well known bottom-up scenario of
hierarchical cluster formation predicted for the real universe.
Concurrently the variance of the velocities grows approximately
exponentially. The power spectrum was monitored at various times and
continued to show scale-free, power-law, behavior. In the strongly
clustered epoch it turns around at small scales to a negative value
supporting the observed structure of clusters of increasing size. It
was shown that a simple analytic representation of the power spectra
captured the main features of the evolution, including the correct
time dependence of the crossover from the linear to nonlinear regime
and the transition from regular to fractal geometry. The underlying
mechanism for the observed behavior may be a process of
{}``hierarchical virialization'' whereby virialized clusters that
have {}``frozen out'' from the cosmic expansion become the new
{}``particles'' for the next virialization sequence. The
measurements were consistent with the self-similar evolution found
in the simulations of Yano and Gouda using the Zeldovich model
\cite{Gouda_powsp}.

In the well clustered regime, a multifractal analysis of the
configuration space was performed. For the RF and Q models, in
contrast with our earlier work with isothermal and waterbag initial
conditions \cite{MRexp,MRGexp}, here a single dominant scaling range
of the partition function $C_{q}$ was found for all values of $q$
considered. Generalized dimensions $D_{q}$ and scaling index
$\tau_{q}$ were evaluated for $-4<q<4$. Robust fractal behavior was
inferred for $q>0$ from the values of $D_{q}$. For the Q model the
gradual decrease in $D_{q}$ for $q>0$ suggests that the system is
also weakly multifractal. The RF model exhibited the greatest
inhomogeneity. It appeared nearly monofractal with $D_{2}=0.5$. In
addition, the two-body correlation function was constructed and also
exhibited a power-law dependence on the displacement. For large
samples, on the order of $10^{5}$ or more particles, the
mathematical relation between the correlation dimension $D_{2}$, the
spectral index $n$ and the correlation function decay exponent
$\gamma$, $D_{2}=-n=1-\gamma,$ was nicely satisfied. For the Q model
at $\tau=13$, $D_{2}=0.63$ over a finite scale range. For
$\mbox{\ensuremath{n>2}}$ there was evidence of hierarchical
clustering whereas for $n<2$ there was a tendency to form a few
large clusters, but in the absence of self-similar, bottom-up,
evolution. Like the Q and RF models, the Hamiltonian version
undergoes rapid hierarchical clustering for a short time span.
However, the scaling behavior was less robust, and the agreement
among the exponents was weaker. In conclusion, one-dimensional
models can exhibit the type of evolution characteristic of the real
universe. Their evolution does reveal fractal behavior, but on a
limited length scale that changes with time. It also shows that
hierarchical structure formation depends sensitively on the initial
conditions. Their three dimensional counterparts are useful for
constraining models of the early universe.

A number of interesting questions about these models still remain.
In forthcoming work we will explore the connection between symmetry
and the various one-dimensional models and show how additional models
analogous to the Q type can be constructed. We will also establish
a connection between the increase in the variance of the velocity
and virialization. While the fractal properties of the over-dense
clusters appear to be well represented in the current work, the geometry
of the under-dense {}``voids'' has not been revealed. This is under
investigation with a greater number of particles per cluster to improve
the resolution. A complementary approach that may retain more information
in the under-dense regions is provided by the numerical integration
of the Vlasov equation. To finally address the conjecture of Balian
and Schaeffer concerning the bifractality of the universe for this
class of models, this must be completed. The approximate analytic
representation of the power spectra introduced in Section (III) suggests
that the low density inter-cluster regions have fractal dimension
unity, but these results must be considered preliminary. A compelling
issue is the connection between cosmologies in different dimensions.
In an earlier work we showed that the RF model was able to crudely
reproduce the time of galaxy formation in a matter-dominated universe
\cite{MRexp}. It would be intriguing to be able to establish a more
precise correspondence.

\ack

The authors benefitted from interactions with Athanasios Batakis,
George Gilbert, Michael Joyce, Michael Kiessling, Igor Prokhorenkov,
Paul Ricker, Patrick Valageas, and Martin White. The authors appreciate
support provided by the Center for Technologies Resources at Texas
Christian University. B. Miller benefitted from the support of the
Research Foundation of TCU, and from the hospitality of MAPMO at Université
d'Orléans during the Spring of 2009.\\

\bibliographystyle{unsrt}
\bibliography{gravbib8}

\end{document}